\documentclass[fleqn,usenatbib]{mnras}
\usepackage{natbib}
\usepackage{amsmath}
\usepackage{graphicx}
\usepackage{txfonts}
\usepackage[switch]{lineno}

\newcommand{\degree}{^\circ}

\usepackage{hyperref}
\usepackage{xargs}                      
\usepackage[pdftex,dvipsnames]{xcolor}  
\usepackage[colorinlistoftodos,prependcaption,textsize=tiny]{todonotes}
\newcommandx{\unsure}[2][1=]{\todo[linecolor=red,backgroundcolor=red!25,bordercolor=red,#1]{#2}}
\newcommandx{\change}[2][1=]{\todo[linecolor=blue,backgroundcolor=blue!25,bordercolor=blue,#1]{#2}}
\newcommandx{\info}[2][1=]{\todo[linecolor=OliveGreen,backgroundcolor=OliveGreen!25,bordercolor=OliveGreen,#1]{#2}}
\newcommandx{\improvement}[2][1=]{\todo[linecolor=Plum,backgroundcolor=Plum!25,bordercolor=Plum,#1]{#2}}
\newcommandx{\thiswillnotshow}[2][1=]{\todo[disable,#1]{#2}}
\title[Holographic slitless spectrograph: focus and resolution]{
A transmission hologram for slitless spectrophotometry on a convergent telescope beam.
1. Focus and resolution.
}
\author[M.~Moniez et al.]{%
M.~Moniez$^{1}$\thanks{E-mail: moniez@lal.in2p3.fr},
J.~Neveu$^{1}$,
S.~Dagoret-Campagne$^{1}$,
Y.~Gentet$^{2}$,
L.~Le Guillou$^{3}$,
\newauthor
The LSST Dark Energy Science Collaboration\\
$^{1}$ Universit\'e Paris-Saclay, CNRS/IN2P3, IJCLab, 91405 Orsay, France\\
$^{2}$ Laboratoire Ultimate Holography, 13 all\'ee d'Androm\`ede 33160 Saint-Aubin, France\\
$^{3}$ Sorbonne Universit\'e, CNRS/IN2P3,
Laboratoire de Physique Nucl\'eaire et de Hautes \'Energies (LPNHE),
75005 Paris, France
}

\date{Accepted 2021 July 09. Received 2021 July 09; in original form 2021 June 28}
\pubyear{2021}
\begin{document}
\label{firstpage}
\pagerange{\pageref{firstpage}--\pageref{lastpage}}
\maketitle
\begin{abstract}
We report in this paper the test of a plane holographic optical element to be used as an aberration-corrected grating for a slitless spectrograph, inserted in a convergent telescope beam.
Our long term objective is the optimisation of a specific hologram to switch the auxiliary telescope imager of the Vera Rubin Observatory into an accurate slitless spectrograph, dedicated to the atmospheric transmission measurement.
We present and discuss here the promising results of tests performed with prototype holograms at the CTIO $0.9\,$m telescope during a run of 17 nights in May-June 2017.
After their on-sky geometrical characterisation, the performances of the holograms as aberration-balanced dispersive optical elements have been established by analysing spectra obtained from spectrophotometric standard stars and narrow-band emitter planetary nebulae.
Thanks to their additional optical function, our holographic disperser prototypes allow to produce significantly better focused spectra within the full visible wavelength domain $[370,1050]\,$nm than a regular grating, which suffers from strong defocusing and aberrations when used in similar conditions. We show that the resolution of our slitless on-axis spectrograph equipped with the hologram approaches its theoretical performance.
While estimating the benefits of an hologram for the spectrum resolution, the roadmap to produce a competitive holographic element for the Vera Rubin Observatory auxiliary telescope has been established.
\end{abstract}


\section{Introduction and context}

The Vera C. Rubin Observatory uses a 8.4 meter diameter telescope, equipped with a 3.2\,Gpixels back-illuminated CCD camera,
which will be devoted to a 10 year south sky survey with 6 wide-band filters {\it ugrizy} \citep{LSSTScienceBook_2009}. One of the objectives is to reach sub-percent photometric precision, which needs a careful calibration procedure \citep{Ingraham_2016}.
Therefore, the main telescope (Simonyi Survey Telescope, hereafter named SST) will be assisted by an auxiliary telescope (AuxTel)
(diameter 1.2~m, $f/18$, scale at focal plane $105\,\mu$m/arcsec), partly devoted to the measurement of the atmospheric transmission \citep{Burke_2010, Burke_2013, Coughlin_2018} which is one of the main photometric systematic sources of uncertainty at the sub-percent level.
To monitor these spatial and temporal transmission variations, a slitless spectrograph is inserted in the
AuxTel converging beam, following the original idea from David Monet (US Naval Observatory, Flagstaff Station) \citep{Stubbs_2021}.
It will allow to obtain the spectra of spectrophotometric standards from $370\,$nm to $1050\,$nm.
This wide wavelength domain is necessary since very different atmospheric features will be studied such as the water vapor absorption band (around $950\,$nm) and aerosol absorption variations in the bluest part of the spectrum. 
The spectrograph disperser is a grating, inserted in a filter wheel, illuminating the on-axis camera with the zero-th and first diffraction orders of the convergent beam. The filter wheel can be moved at a distance between $10$ to $20\,$cm from the sensor (in our case, a Charge-Coupled Device or CCD).
This design allows to switch between photometric and spectrophotometric studies with the same instrument by simply rotating the filter wheel.
The advantages of a slitless spectrograph are the pointing facility and the possibility to make spectrophotometric measurements unaffected by slit vignetting. Achieving this nevertheless requires to model accurately the instrument throughput and its wavelength-dependent point-spread function (PSF). Observation nights at the Cerro Tololo Inter-American Observatory CTIO $0.9\,$m telescope ($f/13.7$, scale at focal plane $60\,\mu$m/arcsec) have been conducted to test the feasibility of this concept and to compare classical periodic gratings with holographic dispersors.

Starting from the observation that a periodic grating presents limitations when used with a convergent beam (Sect. \ref{Sect:ronchi}), we propose a solution based on a holographic optical element.
In Sect. \ref{Sect:producing} we describe the holographic grating principle, and its design and production for its use at the CTIO $0.9\,$m telescope.
A complete set of tests has been performed at this telescope.
Sect. \ref{Sect:telescope} details the geometrical characterisation of the hologram deduced from the systematic scan of point-source pointings around the optical center.
Then the performances are shown in Sect. \ref{Sec:performances}, in terms of focus, spectral resolution, and transmission for the first and second orders of diffraction.
In the discussion (Sect. \ref{Sect:discussion}), we compare the performances of two types of holograms -- amplitude and phase -- with the performances of a Ronchi periodic grating and a blazed periodic grating. We summarise the lessons learned from these first holographic prototypes with a list of requirements for the final hologram adapted to the AuxTel configuration.
We conclude on the benefits of the holographic additional optical function with respect to periodic gratings for the use with a converging beam in Sect.~\ref{Sect:conclusion}.

\section{Limitations of a periodic grating used as a spectrograph with a convergent beam}
\label{Sect:ronchi}
\begin{figure}
\begin{center}
\includegraphics[width=\columnwidth]{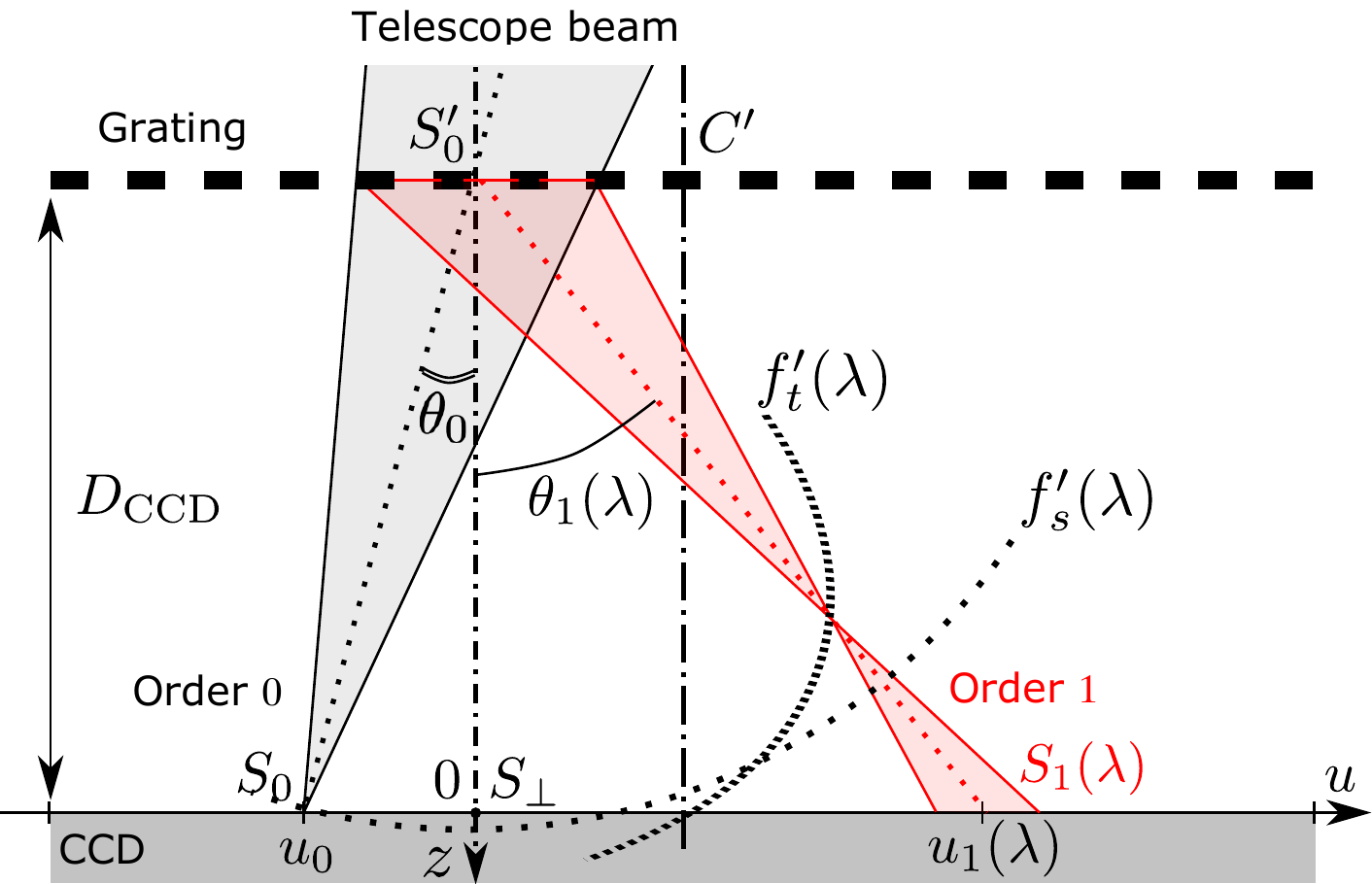}
\end{center}
\caption[]
{Notations used for the dispersion relation and the focusing properties of a regular grating. The $u$ axis is the dispersion axis. The dotted lines marked $f'_t(\lambda)$ and $f'_s(\lambda)$ show the locations of the tangential and sagittal foci of the first-order diffracted image as $\lambda$ varies, for a given $\theta_0$ (which fixes $D_{\mathrm{CCD}}$).
}
\label{fig:dispersion}
\end{figure}

The optical properties of regular periodic gratings used in a convergent beam have been extensively studied in many papers and books (see e.g. \cite{Ives:17,Monk:28,Murty:62,Hall:66,schroeder2000astronomical}) as well as their implementation in slit spectrographs (see e.g.~\cite{Gillieson_1949,Ferraro:00}). Despite the use in a convergent beam instead of a plane wave, these papers show that the Fermat's principle implies that the grating formula is still valid at zero-th order :
\begin{equation}\label{eq:grating}
    \sin \theta_p(\lambda) - \sin \theta_0 = p N_{\rm eff} \lambda,
\end{equation}
where the angles are those of the projection in the plane perpendicular to the grating lines (see Fig.~\ref{fig:dispersion}); $\theta_0$ is the angle of the projected telescope beam axis with respect to the normal to the grating surface, $p$ is the diffraction order, $\theta_p(\lambda)$ is the projected corresponding diffracted angle, and $N_{\rm eff}$ is the effective spatial frequency of grating lines at the position of the central ray of the light beam (hereafter called chief ray).

The other angle characterising the incident beam axis is its angle with respect to the grating lines direction. This angle is identical for all the emergent beams, whatever be the diffraction order \citep{Spencer:62, Harvey2019}.

\begin{figure}
\centering
\includegraphics[width=\columnwidth]{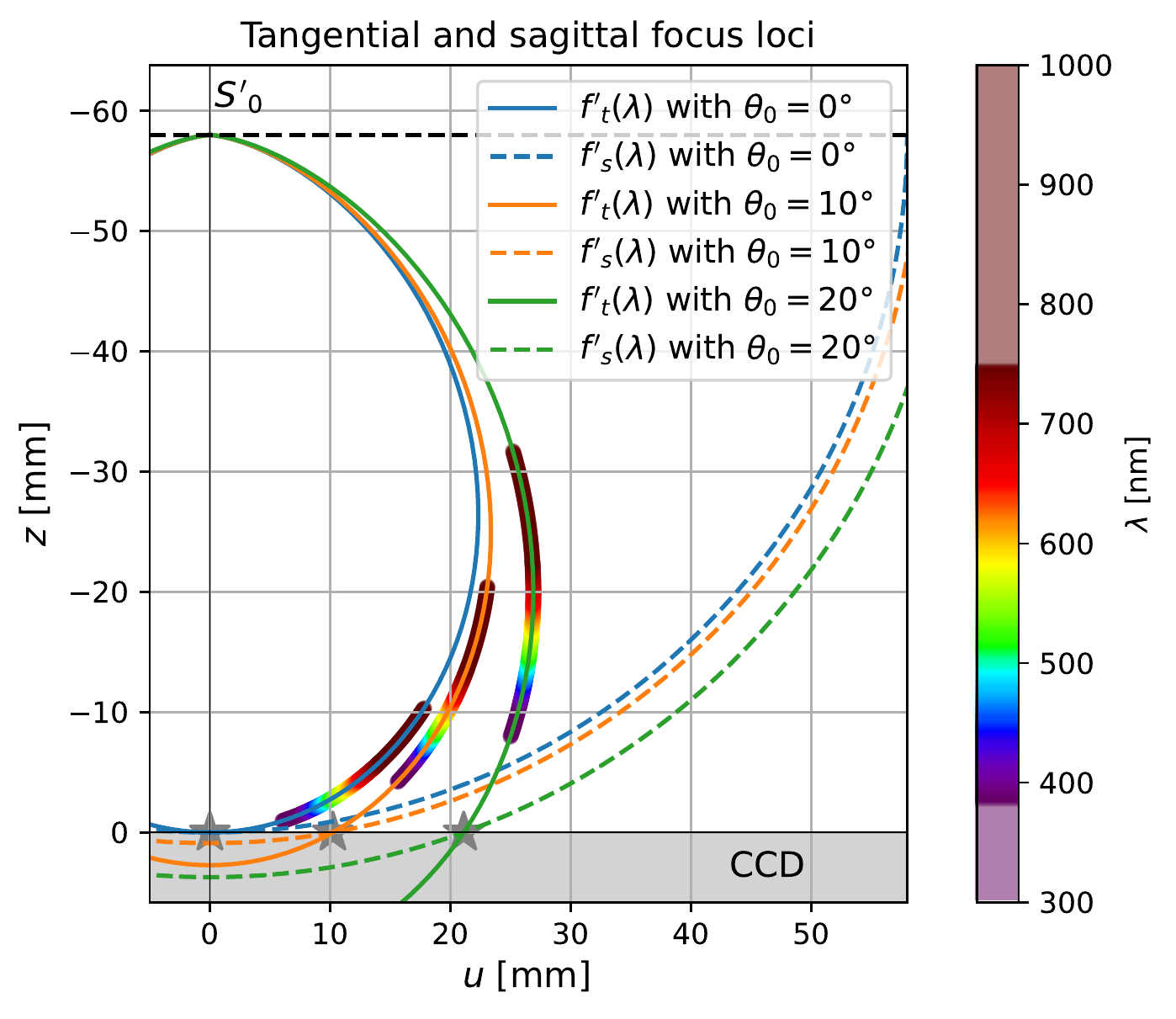}\\
\includegraphics[width=\columnwidth]{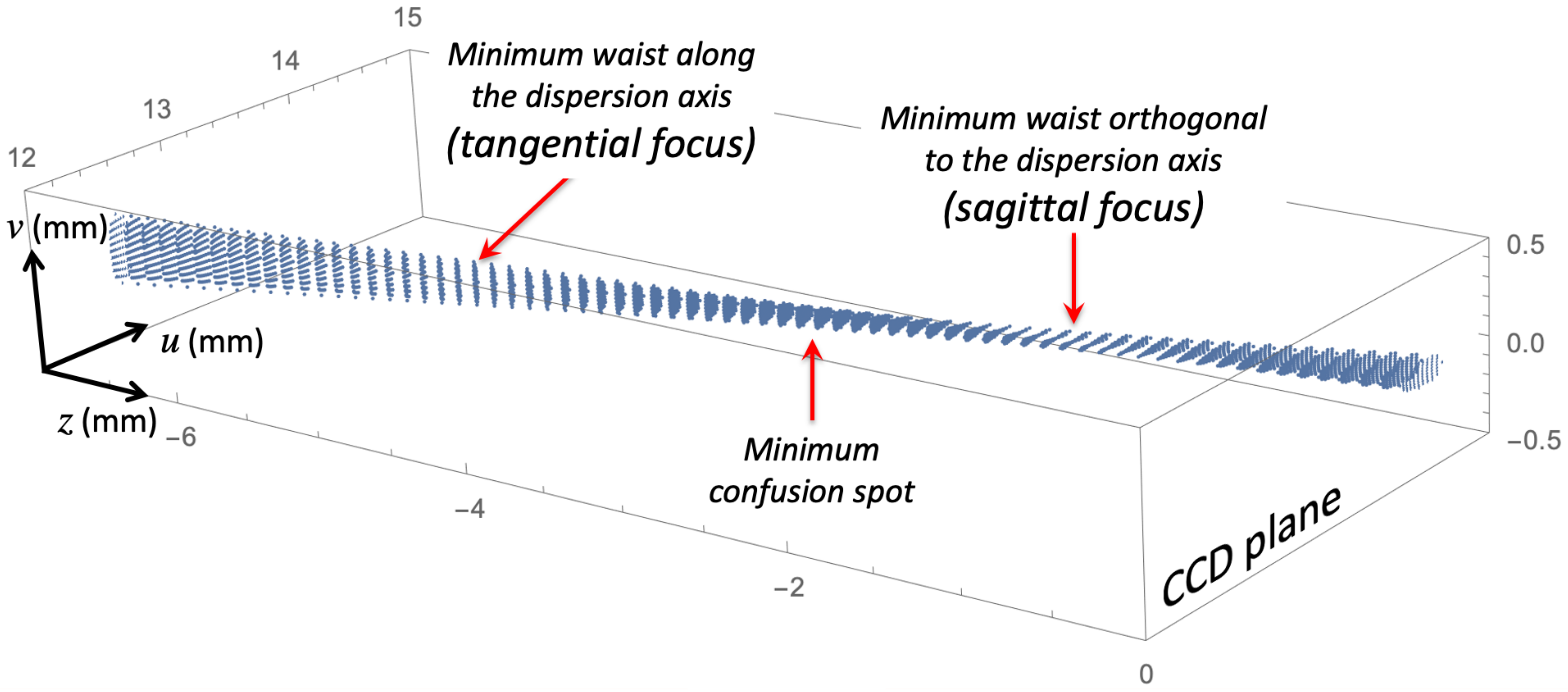}
\caption[]{
{\it Top:} tangential and sagittal focus loci for the first diffraction order with $D_{\mathrm{CCD}} = 58\,$mm and a Ronchi grating with $N_{\mathrm{eff}}=350\,$lines/mm, and different incident beam angles $\theta_0$. The grey stars represent the focused order 0 at $z=0$ for each $\theta_0$ (beam pointing downwards, chief ray passing through $S'_0$). The colored dots indicate the diffraction angles $\theta_1(\lambda)$ for different wavelengths $\lambda$ along the tangential focus locus with respect to the grating normal's direction.

{\it Bottom:} view of the 3D beam structure at $600\,$nm with a Ronchi 400\,lines/mm computed from  a BEAMFOUR ray-tracing simulation, showing the definition of tangential and sagittal foci. (see also Appendix~\ref{sec:psfsim}).}\label{focus2}
\end{figure}

Then aberration terms can be computed at first order (defocusing terms) and second order (coma terms). In particular, in the plane orthogonal to the grating lines (the dispersion plane also called the tangential plane), the tangential focus distance is given by a lemniscate curve (see Fig. \ref{focus2} for the definition of the tangential and sagittal foci):
\begin{equation}
    f'_t(\lambda) = \frac{D_{\mathrm{CCD}}}{\cos\theta_0} \frac{\cos^2 \theta_p(\lambda)}{\cos^2 \theta_0} 
\end{equation}
with $D_{\mathrm{CCD}}=S'_0S_{\perp}$ the distance between the disperser and the CCD (or sensor). In the orthogonal plane, the sagittal focus locus is a circle of radius $f'_s=S'_0S_0=D_{\mathrm{CCD}}/\cos\theta_0$, when the order $0$ is focused on the CCD. An illustration of these formula for a distance $D_{\mathrm{CCD}} =58\,$mm like expected on the CTIO telescope is shown in Fig.~\ref{focus2}.

\begin{figure}
\begin{center}
\includegraphics[width=\columnwidth]{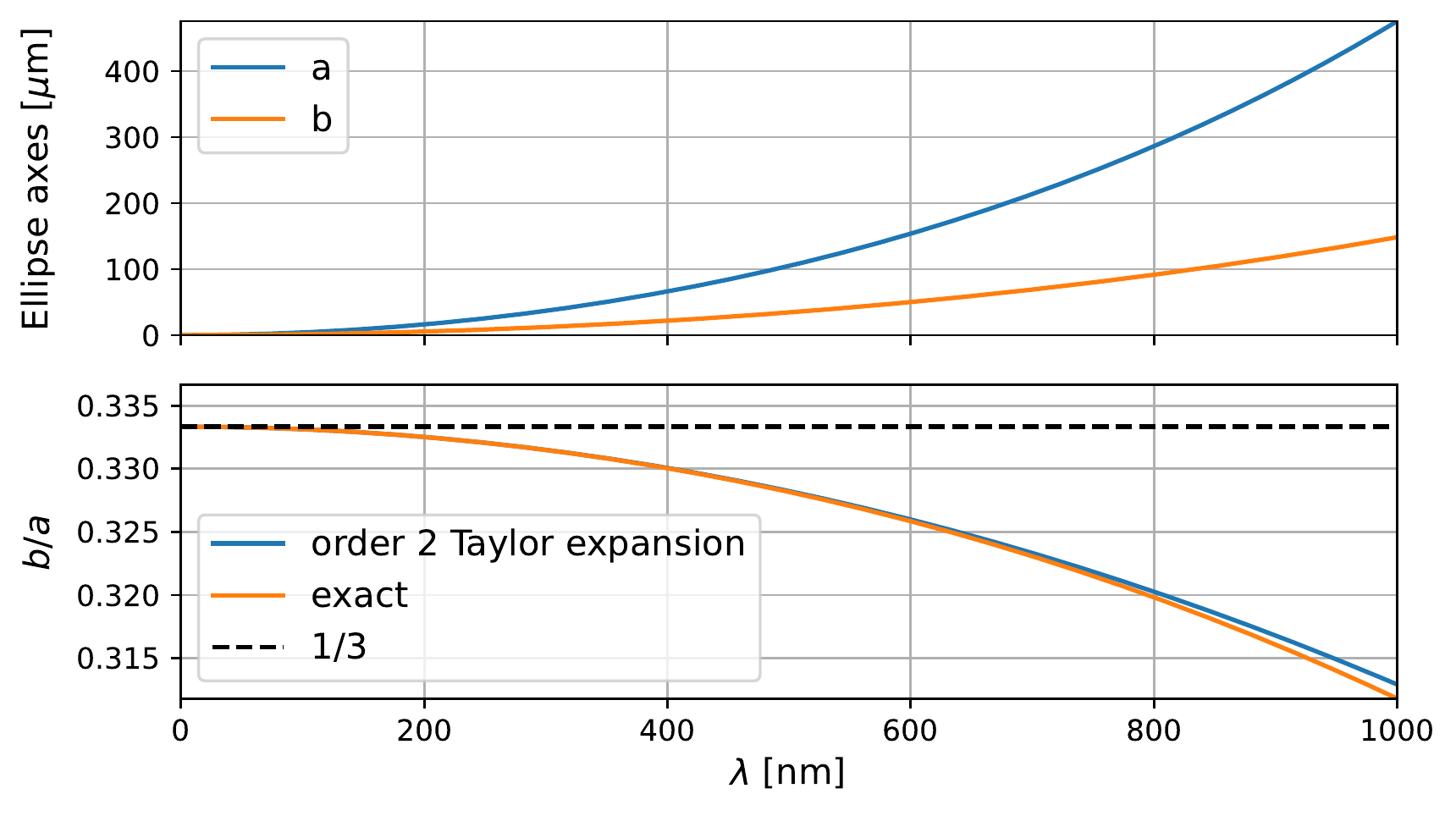}
\end{center}
\caption[]{Ellipse main axes $a$ and $b$ (up) and $b/a$ ratio (down) of the optical PSF for a periodic grating illuminated with a convergent beam at $\theta_0=0$, as a function of wavelength $\lambda$ with $N_{\mathrm{eff}} = 350$\,lines/mm. The second order Taylor expansion superimposes very well with the exact computation.}
\label{eccentricity}
\end{figure}

Therefore, the defocusing increases with the wavelength $\lambda$ and $N_{\mathrm{eff}}$.
The monochromatic beam structure near focus can be characterized from the disperser to the CCD by i) the position of the minimal waist within the dispersion plane, ii) the minimal confusion spot, and iii) the position of the minimal waist orthogonally to the dispersion plane (in the sagittal plane). The effect that will limit the wavelength resolution when extracting the spectrum will be the extension of the spot within the dispersion plane at the detector position, as shown in Fig.~\ref{focus2} (bottom), established with a BEAMFOUR simulation \citep{Beamfour_2016}.

\begin{figure*}
\begin{center}
\includegraphics[width=\textwidth]{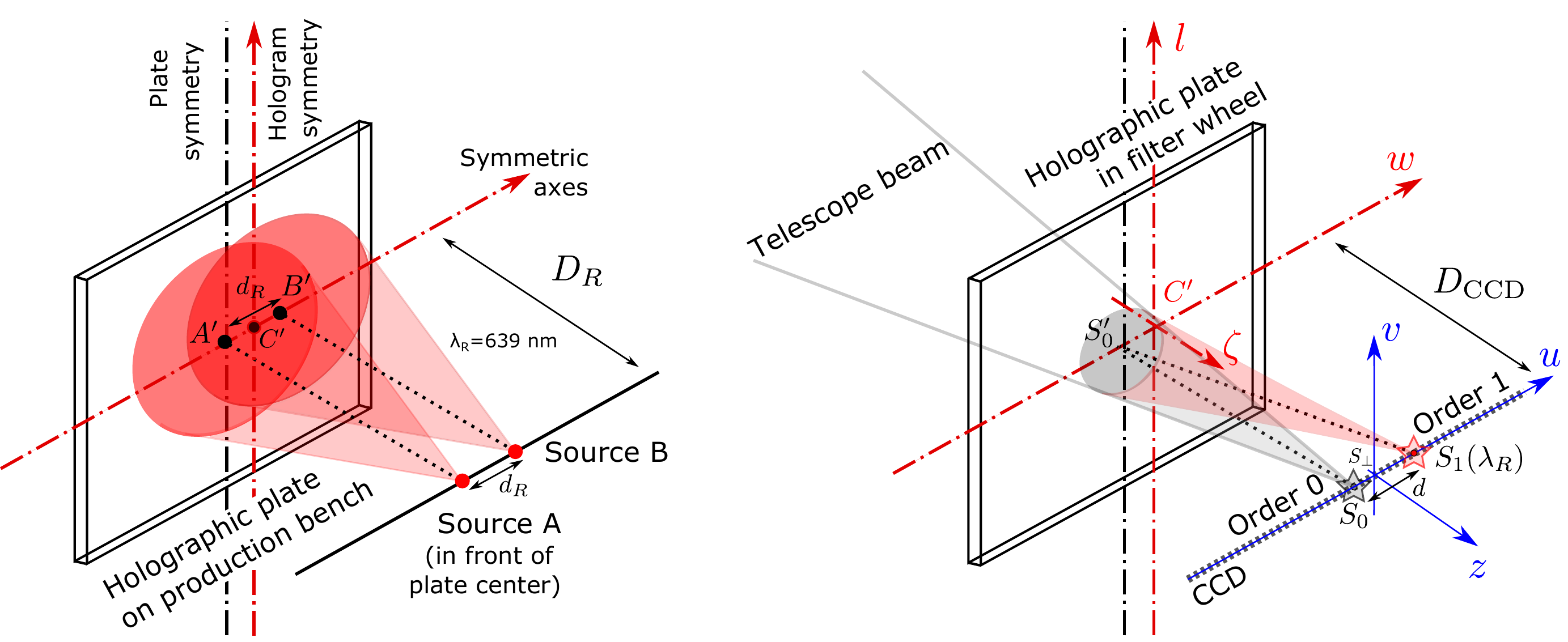}
\end{center}
\caption[] {
\textit{Left: recording of the hologram.} The interference pattern of two illuminating point sources A (reference wave) and B (image wave), produced by a split laser beam  of wavelength $\lambda_R=639\,$nm, is recorded on the holographic sensitive plate emulsion.
The two sources are placed at the distance $D_R$ from the plate and spaced by a distance $d_R$.
\textit{Right: reading of the hologram.} The telescope beam, converging at the reference wave point source position $A$, is diffracted by the hologram. By construction, if $D_\mathrm{CCD}=D_R$, the zero-th order of the observed star is focused at the place where was source $A$, and the first order at wavelength $\lambda_R$ is focused at the place where was source $B$, {\it i.e.} $d=d_R$. The other first order wavelengths are also well focused on the CCD plane. In both figures, red dotted lines give the symmetric axis of the interference pattern, and the black dotted line gives the glass plate median axis. The coordinate frames of the disperser $(w,l,\zeta)$ and of the spectrum $(u,v,z)$ are represented.}
\label{prod-holo}
\end{figure*}

At first order, the defocus produces elliptic patterns on the CCD for point sources if the telescope focus has been tuned on the zero-th diffraction order. The major axis $a$ (along the dispersion direction) and the minor axis $b$ (perpendicular to the dispersion direction) of the ellipses can be computed at the intersection of the beam with the CCD plane:
\begin{align}
     a & = \frac{D}{2} \left(\frac{\cos^3 \theta_0}{\cos^3 \theta_p(\lambda)}-1\right), \qquad
     b  = \frac{D}{2} \left(\frac{\cos\theta_0}{\cos \theta_p(\lambda)}-1\right) 
\end{align}
with $D$ the telescope diameter. After a second order Taylor development assuming small $\theta(\lambda)$ diffraction angles and $\theta_0\ll 1$, the $b/a$ ratio is:
\begin{align}
    \frac{b}{a} & \approx \frac{1}{3}\left[ 1 - \cfrac{1}{2}\left(N_{\mathrm{eff}}\lambda\right)^2\right] \approx \frac{1}{3}
\end{align}
giving an ellipse eccentricity $e\approx \sqrt{8}/3$. The dependency with the wavelength is rather small (see Fig.~\ref{eccentricity}) and even the zero-th order development is a very good approximation of the elliptic PSF pattern.
These formula show also that the optical PSF of such a slitless spectrograph can be modelled (see also Appendix~\ref{sec:psfsim}).
When designing a system using a regular grating with this type of configuration, one can consider to minimize the impact of defocusing, for example by reducing the wavelength range, or the dispersion power, or by conceding zero-th order defocusing to improve first order focusing.

For instance, for the CTIO $0.9\,$m telescope configuration of Fig. \ref{defocus} where a Ronchi with 400 lines/mm is installed at $D_{\mathrm{CCD}} \approx 58\,$mm from the CCD plane,
the defocusing in the deflection plane reaches $8.7\,$mm in red ($800\,$nm) and $13.4\,$mm in infra-red ($1000\,$nm), enlarging the image spot FWHM to $\sim 0.5\,$mm ($\sim 8\,$arcsec) and $\sim 1.0\,$mm ($\sim 16\,$arcsec) in the dispersion direction on the CCD plane.
For AuxTel, the situation will be less degraded; nevertheless, with a periodic grating of 150 lines/mm installed at $D_{\mathrm{CCD}} \approx 200\,$mm, the defocusing will be $\sim 3.8\,$mm in red and $\sim 6.7\,$mm in infra-red, enlarging
the image spot FWHM to $\sim 0.15\,$mm ($\sim 1.5\,$arcsec) and $\sim 0.3\,$mm ($\sim 3.0\,$arcsec) in the dispersion direction.

In addition to the defocusing, optical geometric aberrations like coma aberrations are expected on the CCD plane, affecting the light repartition in the elongated spot (refer to Appendix~\ref{sec:psfsim} showing BEAMFOUR simulations).
However, in \cite{Murty:62} it has been shown that the coma aberration can be reduced or even cancelled with a grating using a varying ruling width that can be obtained by the interference of two coherent point sources. These kind of holographic dispersers have been extensively studied in many theoretical papers (see e.g.~\cite{Murty:71,Noda:74,Noda_sim:74, Hutley_1976,Hettrick:84,Vila_88,Palmer:89, SINGH2000401,Palmer_2000, Goodman_2017}). Holographic gratings were also widely experimentally tested (see e.g.~\cite{LABEYRIE19695,Rudolph1967,SCHMAHL1977195,Namioka_1976}) but mostly with concave surfaces. This leads us to the conception of a plane holographic element to design a simple slitless spectrograph with good focus and optical distortions properties, allowing to convert a telescope into a wide wavelength range spectrophotometer.

\section{Description and production of the holographic optical element}
\label{Sect:producing}

\subsection{Principle}
Our goal is to produce an hologram that forces the diffracted spectrum of a converging beam to be correctly focused on the sensor plane. 
Such an hologram, designed to convert a telescope equipped with a CCD camera imager into a spectrophotometric instrument, is specific to the geometry of each telescope, because it depends on the distance $D_{\mathrm{CCD}}$ between the plane of the hologram (usually inserted within a filter wheel), the sensor size, and the desired dispersion power.

The production of an hologram is illustrated Fig. \ref{prod-holo}. One has to record
the interference pattern of two spherical waves at a reference wavelength $\lambda_R$, issued from two coherent point-sources $A$ (reference wave) and $B$ (image wave). The $A$ and $B$ sources are positioned respectively at the expected telescope beam focus point (order 0) and at the requested first order diffracted image for $\lambda_R$, ideally both at a distance $D_R = D_\mathrm{CCD}$ from the holographic plate.

According to the holography principles, when the hologram is back illuminated with the reverse reference wave converging in $A$ (order 0 in Fig.~\ref{prod-holo}-right) at wavelength $\lambda_R$, it diffracts a spherical wave converging at position $B$ (order 1) and $d=S_0S_1(\lambda_R)=AB=d_R$ if $D_\mathrm{CCD}=D_R$. If the reference wave is replaced by a telescope beam converging on $A$ issued from an astrophysical object, then a point-like image at $\lambda_R$ is produced at $B$ (order 1). For wavelengths around $\lambda_R$, by continuity, the diffracted wave is also focused near the line $(AB)$, which is the optical function that we are looking for to produce a correctly focused spectrum from a convergent beam in the visible. This characteristic will be quantified throughout this paper. Note that, conversely, the $-1$-th order of the spectrum is defocused.

\begin{figure}
\begin{center}
\includegraphics[width=8.cm]{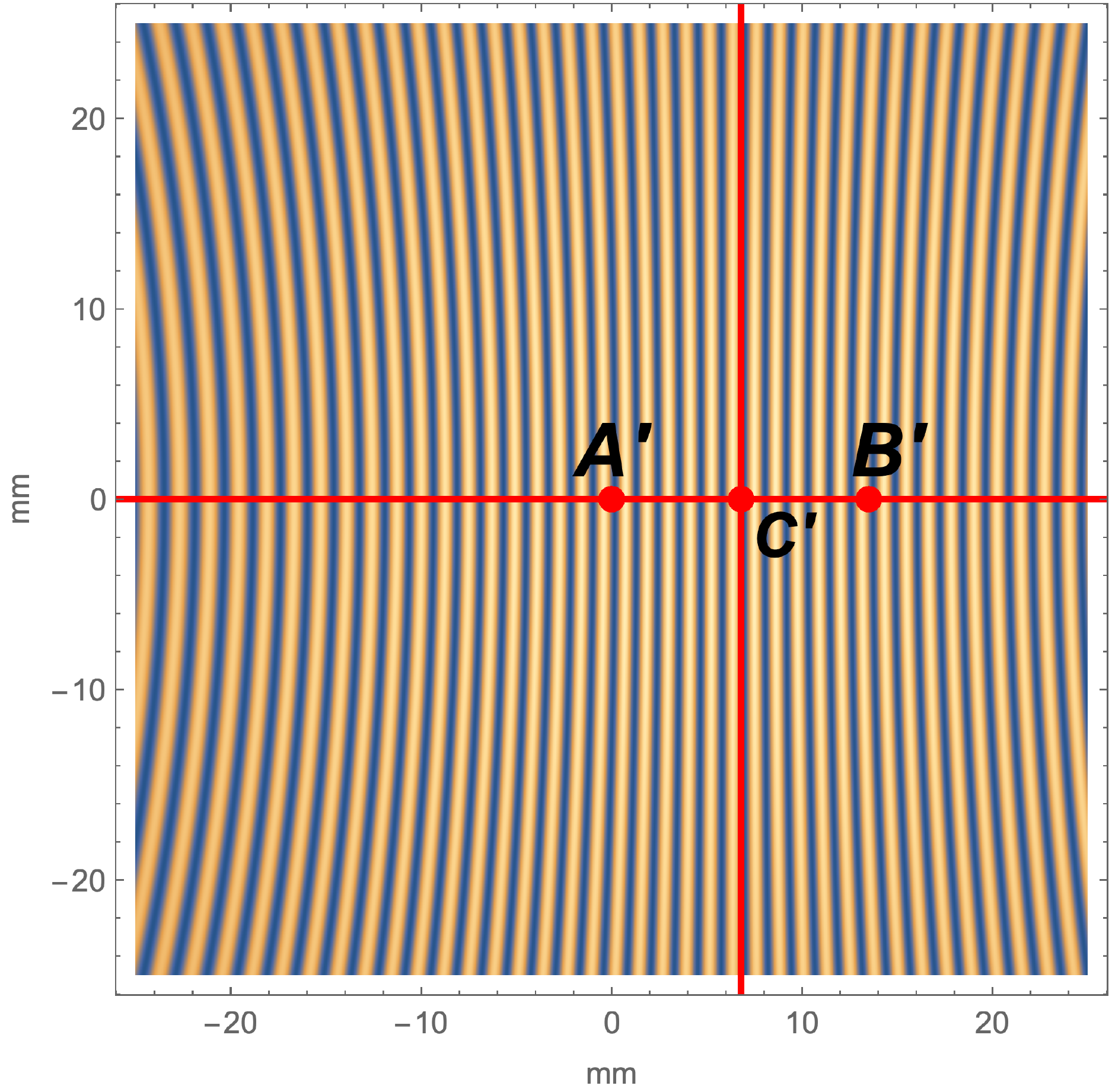}
\end{center}
\caption[]{The intensity interference pattern recorded on the $5\,\mathrm{cm}\times 5\,\mathrm{cm}$ hologram are confocal hyperboloids, as shown here by a simulation corresponding to the holograms produced for this study ;
here only 1 line every 400 is represented (and zoomed).
Note that the symmetry center C' at the intersection of symmetry axis (red lines) does not coincide with the optical center (A' projection of A, perpendicular to the sensor plane).}\label{pattern}
\end{figure}

The recorded interference pattern is made of confocal hyperboloids (Fig.~\ref{pattern}). It is not invariant by translation unlike usual regular gratings, and has two symmetry axes (the $(A'B')$ line and the orthogonal line passing through $C'$ the center of the $[A'B']$ segment, where $A'$ and $B'$ are the orthogonal projections of $A$ and $B$ on the plate). The spatial frequency of the lines decreases from the center $C'$ to the edges of the hologram; as a consequence, the left and right edges of the $4.1\,$mm diameter beam of the CTIO telescope entering the hologram (grey circle on Fig. \ref{prod-holo}-right) are unequally diffracted. The result is an optical focusing function in addition to the dispersion function. As a consequence, the optical center $A'$ of this element (different than the symmetry center $C'$) needs to be aligned with the position $A$ of the undeflected (zero order) beam.
To get this corrective optical function we built a specifically designed optical bench to produce the hologram. The bench's geometry depends only on the distance to the focal plane $D_R=D_{\mathrm{CCD}}$ and on the requested dispersion power (defined by the distance $d_R=AB$).

\subsection{Dispersion and focus properties}

\begin{figure}
\centering
\includegraphics[width=\columnwidth]{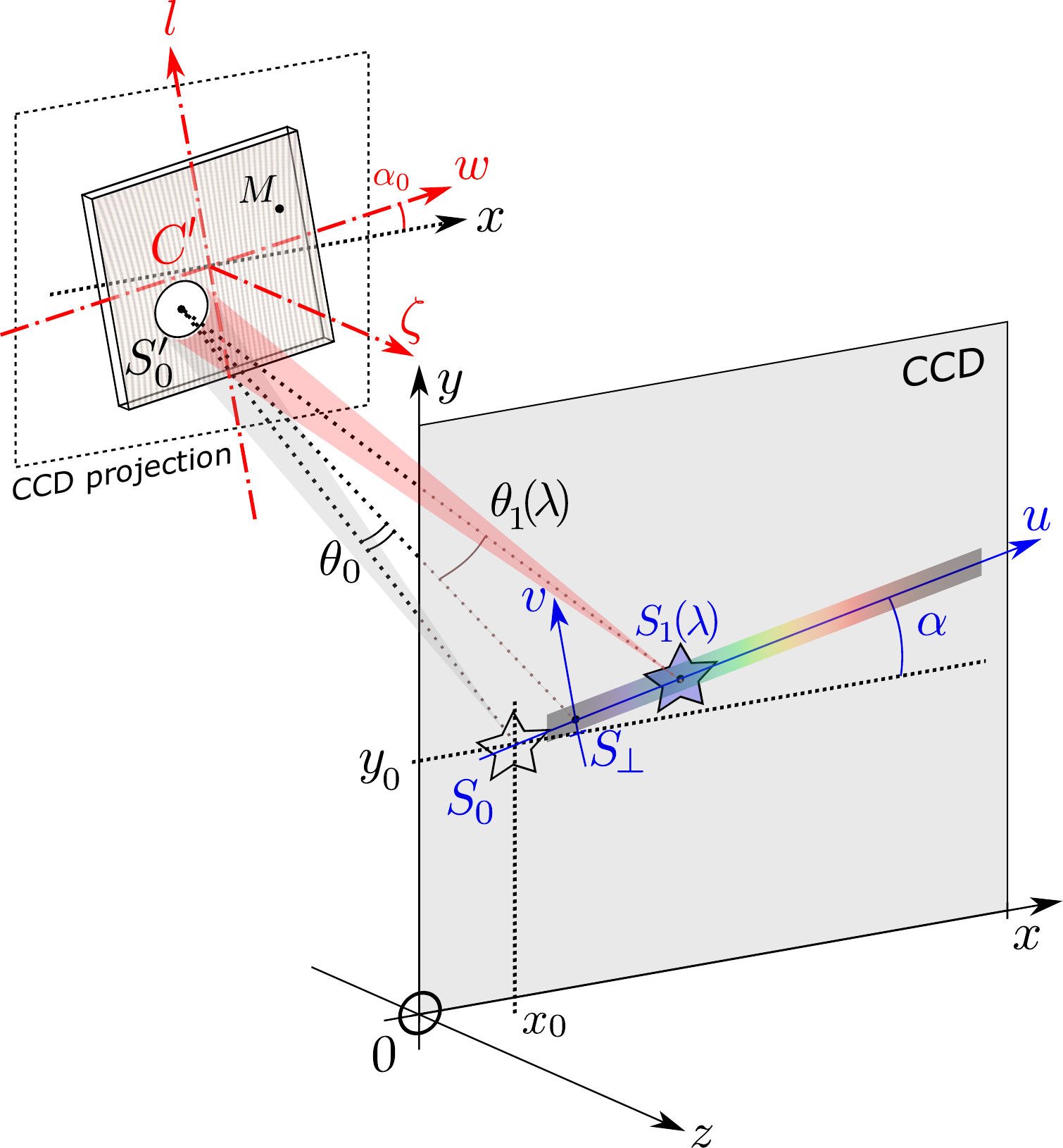}
\caption[]{
Coordinate systems for the description of the dispersion and focusing properties of holographic gratings. The points $S_0$ and $S_1(\lambda)$ are respectively the centroids of the order 0 and order 1 on the CCD at wavelength $\lambda$, while $S'_0$ is the incident point of the chief ray on the hologram. The angle $\alpha_0$ is the orientation angle of the hologram with respect to the $x$ axis, while $\alpha$, depending on the position of $S'_0$,
is the dispersion axis angle with respect to the same $x$ axis. The $(x,y), (u,v), (w,l)$ frames are all coplanar but have different orientations and origins.}
\label{fig:frame}
\end{figure}

A complete description of the dispersion and focusing properties of holographic gratings is developed in \cite{Noda:74}, using the Fermat principle for any hologram ({\it i.e.} any position of the two coherent sources) and any position of the order 0. Hereafter we propose a rewriting of \cite{Noda:74} results using $S'_0$, the incident point of the chief ray on the hologram, as the reference point to apply the Fermat principle. We use the coordinate system $(u,v,z)$ with origin $S_\perp(0,0,0)$ where $u$ is the dispersion axis, the suitable frame to describe the properties of the hologram from the record of the diffraction orders on the CCD (see Fig.~\ref{fig:frame}).

Figure \ref{fig:frame} shows the direct and first order diffracted light-rays issued from the central ray (chief ray of a conical beam) entering the hologram at the position $S'_0(w'_0,l'_0,0)_{w,l,\zeta} = S'_0(0,0,-D_\mathrm{CCD})_{u,v,z}$
and arriving at $S_0(u_0,0,0)$ and $S_1(u_1(\lambda),0,0)$.
We consider another light-ray from the conical beam converging at point $S_0$, crossing the grating surface at a point $M(u,v,-D_\mathrm{CCD})$. Let $n(S'_0,M)$ be the (fractional) number of grooves from $S'_0$ to $M$.
The condition to obtain a constructive interference at position $S_p$ (perfect imaging for order $p$) between the light diffracted at $M$ and the light diffracted at $S'_0$ can be expressed by the stationarity of the light-path function $F$ defined as
\footnote{Here, the $\lambda$ dependencies are implicit to facilitate the reading.}:

\begin{equation}
F(M) = (MS_p - MS_0) - (S'_0S_p - S'_0S_0) + p\lambda n(S'_0,M)
\label{light-path}
\end{equation}
adapted from \cite{Noda:74} for transmission holograms, where:
\begin{align}
MS_p & = \sqrt{(u-u_p)^2+v^2+D_\mathrm{CCD}^2} \notag \\
MS_0 & = \sqrt{(u-u_0)^2+v^2+D_\mathrm{CCD}^2} \notag\\
S'_0S_p & = \sqrt{u_p^2+D_\mathrm{CCD}^2} = r_p \notag\\
S'_0S_0 & = \sqrt{u_0^2+D_\mathrm{CCD}^2} = r_0. \notag
\end{align}
The position $S_p$ satisfying this stationarity condition depends on $\lambda$ and for $p=1$ it is represented as $S_1(\lambda)$ on Fig. \ref{fig:frame}.

What is the meaning of the explicit $\lambda$ term in $F(M)$ ? Since wave fronts are defined as surfaces of constant phase, the optical path difference between the principal path $(S_0S'_0S_p)$ and the general diffracted path $(S_0MS_p)$ need
to be an integer multiple of the diffracted wavelength \citep{Palmer:89}.
The light-path difference $pn(S'_0,M)\lambda$
from $S'_0$ and $M$ states that the phase shift for a ray passing through $M$ is the same as for the ray passing through $S'_0$  (modulo $2\pi$).
For instance, in the case of a regular grating of spatial frequency $N_{\rm eff}$ with straight grooves orthogonal to the $w$ axis, we have simply $n(S'_0,M) = w.N_{\rm eff}$.
For an hologram recorded from two coherent sources $A(u_A,v_A,z_A)$ and $B(u_B,v_B,z_B)$, the number of grooves between $S'_0$ and $M$ is given by the light-path difference:
\begin{equation}
n(S'_0,M) = \left[(MB - MA) - (S'_0B - S'_0A)\right]/\lambda_R.
\end{equation}
Note that $z_A$ and $z_B$ are not necessarily equal in the plane $z=0$ because the CCD can be tilted and at a different distance from the holographic emulsion than the sources were when the hologram was recorded.

Appendix~\ref{app:fermat} develops the computing of the power-series expansion of
$F(M)$ and the consequences of its stationarity on the diffracted beam. Below we summarize the main conclusions.
\begin{itemize}
    \item
    The dispersion axis, characterized by the angle $\alpha(w'_0,l'_0)-\alpha_0$, is locally orthogonal to the mean orientation of the grooves at $S'_0$.
    \item
    The grating formula (\ref{eq:grating}) applies for holographic gratings replacing $N_{\mathrm{eff}}$ by a local $N_{\mathrm{eff}}(w'_0,l'_0)$ line density varying with the incident point $S'_0$ of the chief ray.
    \item
    The tangential and sagittal focuses as a function of $\lambda$ are given by:
\begin{equation}
f'_t(w'_0,l'_0,\lambda) =\dfrac{D_{\mathrm{CCD}} \cos^2\theta_p(\lambda)}{\cos^3\theta_0 -  p\lambda  D_{\mathrm{CCD}}\left.\dfrac{\partial^2 n}{\partial u^2}\right\vert_{S'_0}}
\end{equation}
\begin{equation}
f'_s(w'_0,l'_0,\lambda) =\dfrac{D_{\mathrm{CCD}}}{\cos\theta_0 -  p\lambda  D_{\mathrm{CCD}} \left.\dfrac{\partial^2 n}{\partial v^2}\right\vert_{S'_0}}
\end{equation}
\end{itemize}
where the partial derivatives are explicitly computed at the end of Appendix~\ref{app:fermat}.

\begin{figure}
\centering
\includegraphics[width=\columnwidth]{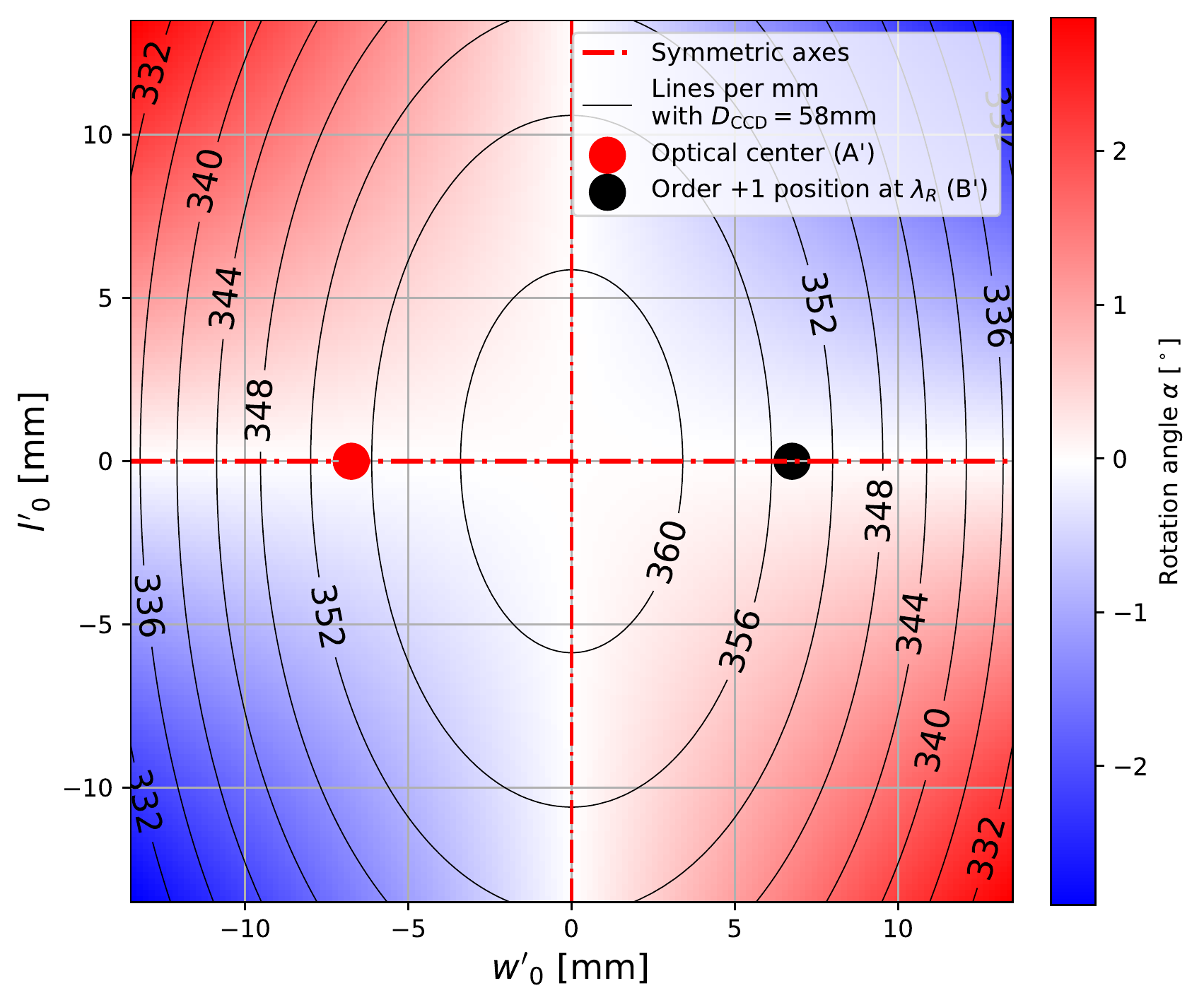}
\caption[]{
Maps of  the groove density $N_\mathrm{eff}(w'_0,l'_0)$ (black contours) and dispersion axis angles $\alpha(w'_0,l'_0)-\alpha_0$ (colored map) for an holographic grating with $D_R=58\,$mm, $d_R = 13.5\,$mm in the $(w,l)$ frame.}\label{fig:Neff_alpha}
\end{figure}

\begin{figure}
\centering
\includegraphics[width=\columnwidth]{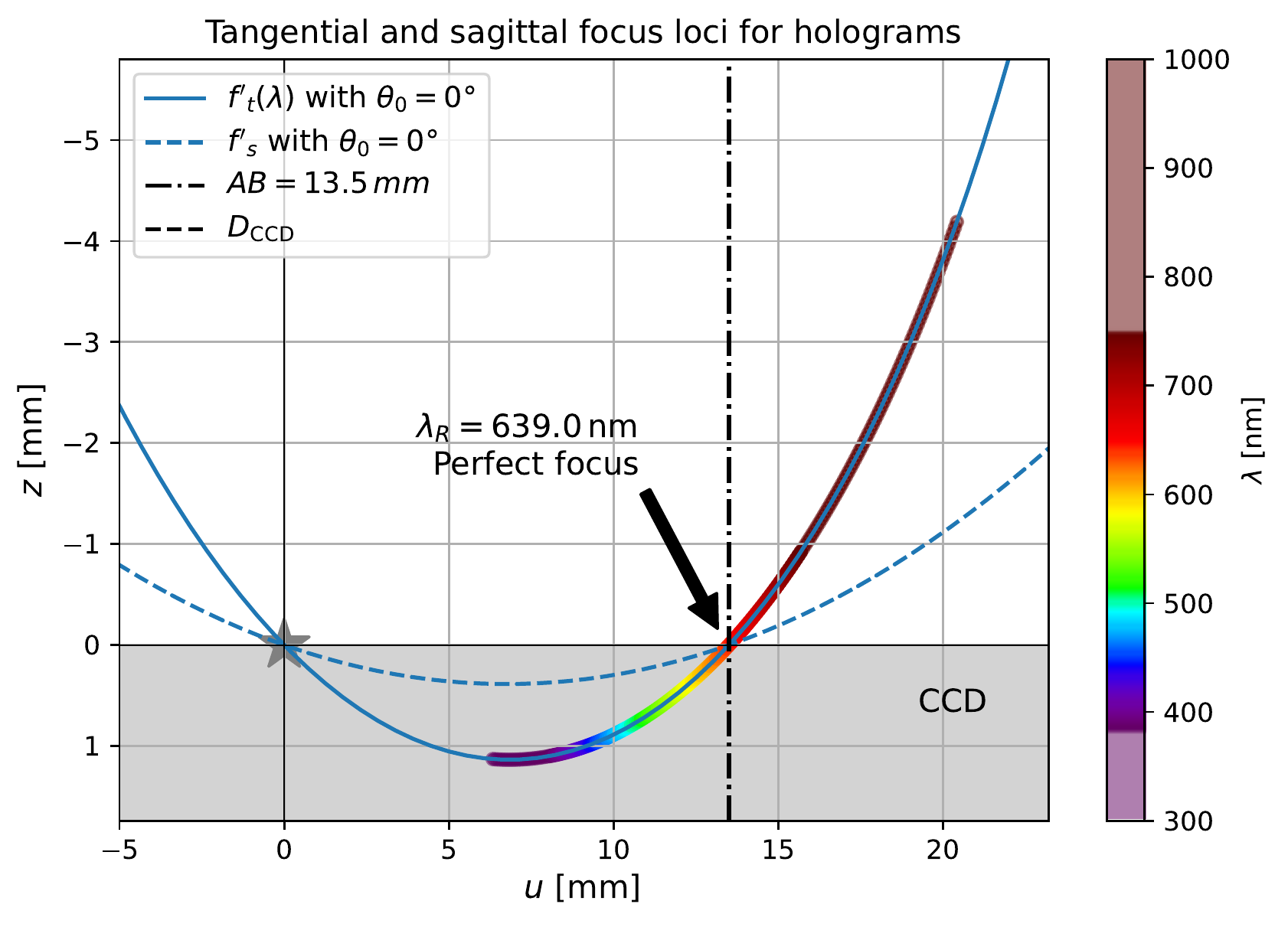}
\caption[]{
Tangential and sagittal focus loci of the first diffraction order for a collinear converging beam at the optical center $w'_0=-d_R/2$ and $l'_0=0$, with $D_{\mathrm{CCD}} = D_R=58\,$mm and  $N_{\mathrm{eff}}=350\,$lines/mm at the optical center. The grey star represents the focused order 0 at $\zeta=D_{\mathrm{CCD}}=D_R$ for $\theta_0=0$. The colored dots indicate the image positions for different wavelengths $\lambda$ along the tangential focus locus.
}\label{focus_hologram}
\end{figure}

\begin{figure}
\begin{center}
\includegraphics[width=\columnwidth]{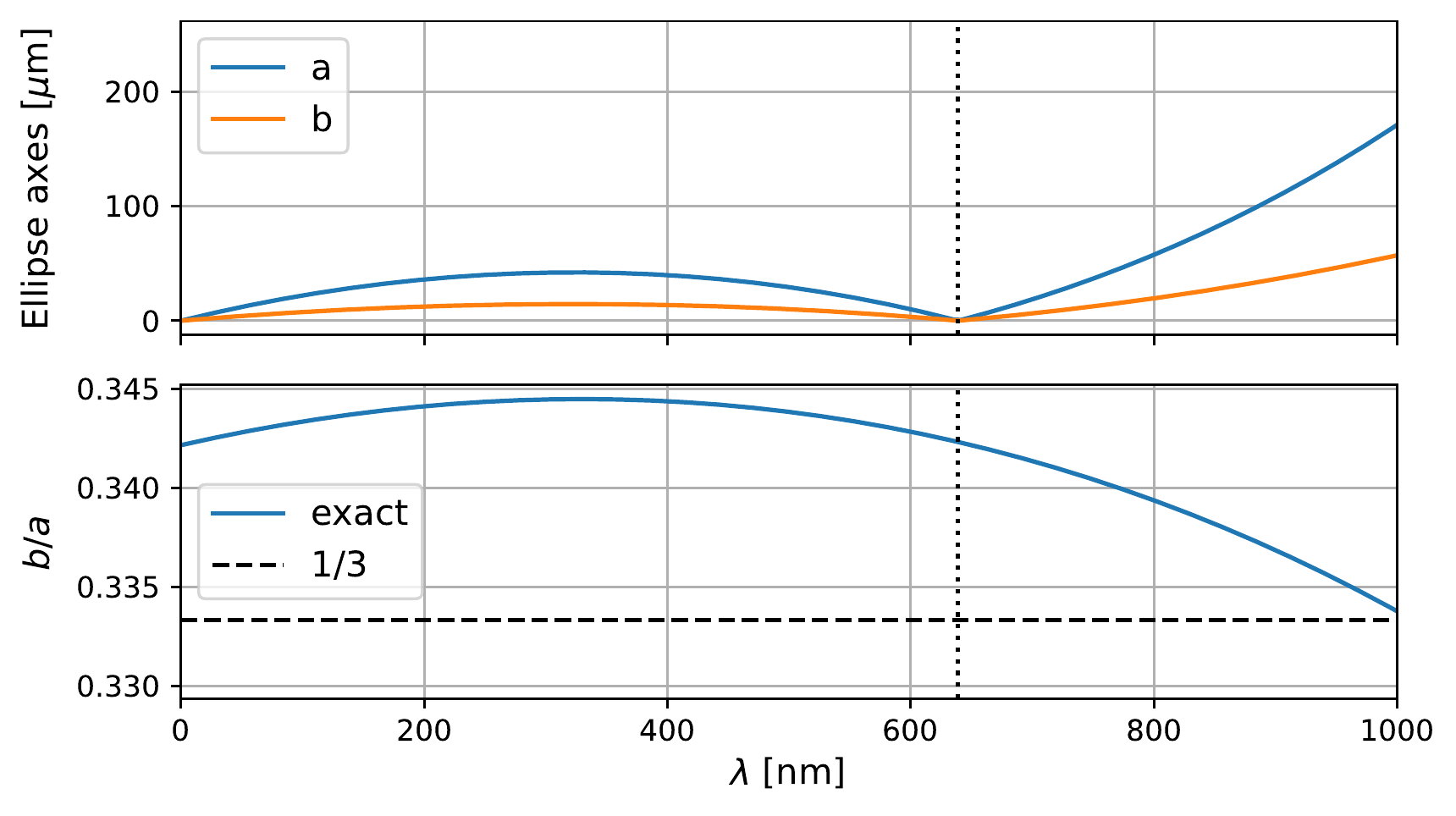}
\end{center}
\caption[]{Ellipse main axes $a$ and $b$ (up) and $b/a$ ratio (down) of the optical PSF for an hologram illuminated with a convergent beam at $\theta_0=0$, as a function of wavelength $\lambda$ with $N_{\mathrm{eff}} = 350$\,lines/mm at the optical center A' and $D_\mathrm{CCD}=D_R=58\,$mm.}
\label{eccentricity_hologram}
\end{figure}

\begin{figure}
\begin{center}
\includegraphics[width=\columnwidth]{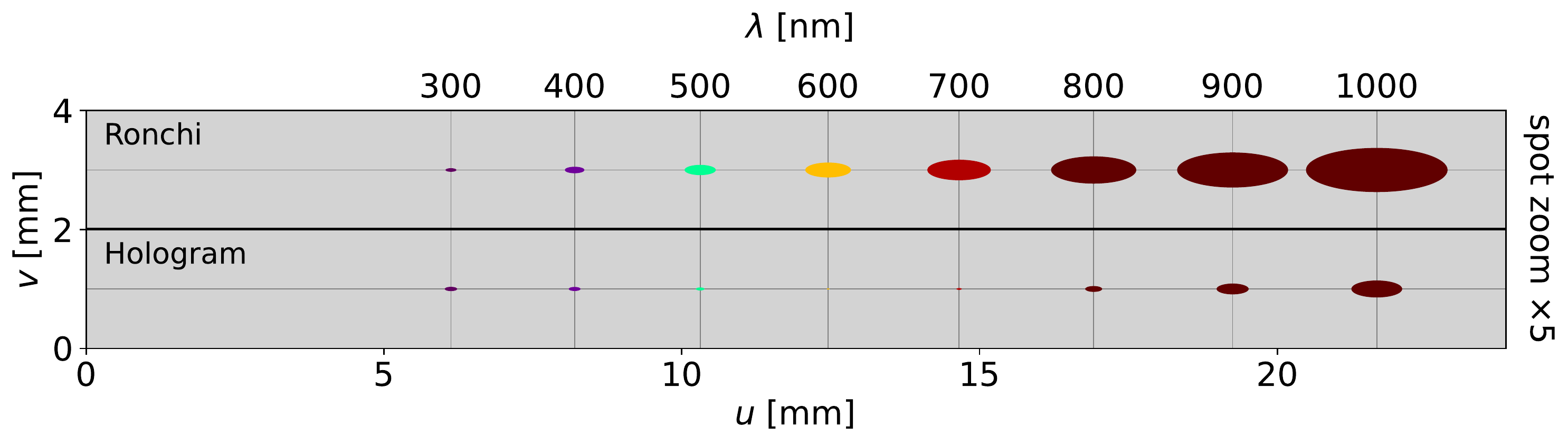}
\end{center}
\caption[]{First diffraction order PSF patterns of the slitless spectrograph for a Ronchi grating (top) and an holographic grating (bottom), both with $N_{\mathrm{eff}} = 350$\,lines/mm at the optical center $A'$ and $D_\mathrm{CCD}=D_R=58\,$mm, for 100~nm-spaced wavelengths from 300 to 1000~nm.}
\label{fig:ellipses}
\end{figure}

The maps of the groove density $N_\mathrm{eff}(w'_0,l'_0)$ and dispersion axis angles $\alpha(w'_0,l'_0)-\alpha_0$ are computed in Fig.~\ref{fig:Neff_alpha}.
We can see that the groove density decreases from the hologram symmetry centre $C'$ and that the latter point is also a saddle point for the dispersion axis angle. 
Concerning the focusing properties, in Fig.~\ref{focus_hologram} we represent the tangential and sagittal loci for holograms designed for the CTIO configuration, when the incident beam is normal ($\theta_0=0$) with $S'_0\approx A'$. Comparing with Fig.~\ref{focus2}, the tangential focus for the wavelength range $\left[300\,\text{nm}, 1000\,\text{nm}\right]$ spans only $\approx 5\,$mm along the $z$ axis while it covers $\approx 10\,$mm for a regular grating with the same $N_{\rm eff}$; moreover, even if the focus is not set on the 0th order but around 600\,nm of the first order, the defocusing effect is still stronger for a regular grating than for an hologram with the same dispersion power.  The first order diffraction images on the sensor plane are ellipses with sizes shown in Fig.~\ref{eccentricity_hologram}: we observe a perfect focus for $\lambda = \lambda_R$ as $a(\lambda_R)=b(\lambda_R)=0$. For any wavelength, values of eccentricities and $b/a$ ratios are close to the ones obtained for regular gratings. Compared with Fig.~\ref{eccentricity}, for the CTIO configuration we expect a strong reduction of the spectrograph PSF size at all visible and infrared wavelengths, as represented in Fig.~\ref{fig:ellipses} and Fig.~\ref{fig:psfsim}.

\subsection{First prototypes}

\begin{figure}
\begin{center}
\includegraphics[width=\columnwidth]{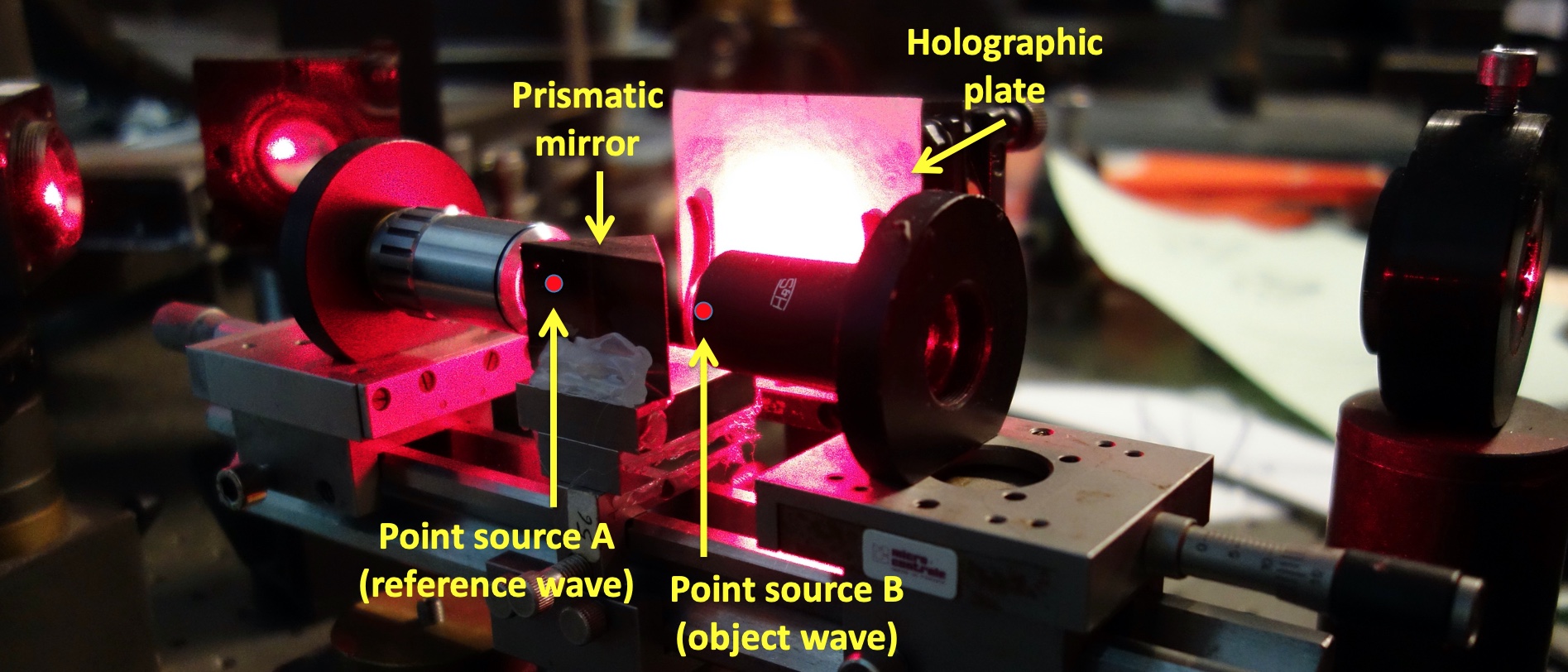}
\includegraphics[width=\columnwidth]{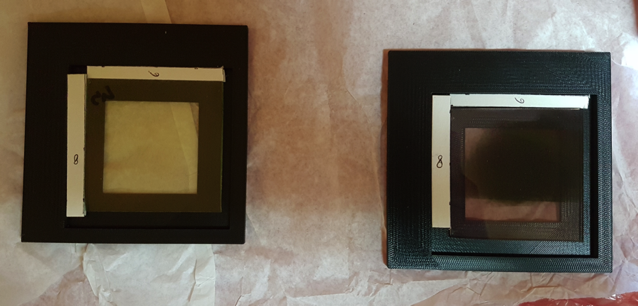}
\end{center}
\caption[] 
{
\textit{Top :} Picture of the holographic production bench at Ultimate Holography (Bordeaux, France)  with sources A and B (red dots indicated by the arrows), deported by a prismatic mirror.
\textit{Bottom:}  phase (left) and amplitude (right) holograms mounted on frames to align their optical centers within our chosen quarter of the CTIO $0.9\,$m telescope CCD camera. 
}
\label{photo-holo}
\end{figure}

The holograms were made at \textit{Ultimate holography}, a company directed by one of the authors (Y. Gentet), based in Bordeaux, France. Two prototypes have been produced dedicated to the CTIO $0.9\,$m telescope configuration ($D_{\mathrm{CCD}}\sim 58\,$mm), using a $\lambda_R=639\,$nm wavelength stabilized laser (Fig. \ref{photo-holo}) :
\begin{itemize}
\item
a silver halide emulsion amplitude hologram, where the interference
pattern is recorded as a high resolution absorption modulation.
\item
a silver halide emulsion phase hologram, where the interference
pattern is recorded as a high resolution optical index modulation.
\end{itemize}

\section{On-sky geometrical characterisation of the holograms}
\label{Sect:telescope}

From May 27th to June 14th 2017, the two silver halide-based holograms have been evaluated together with a Ronchi grating (400 lines/mm)  (hereafter called Ronchi400) and a Thorlabs blazed grating (300 lines/mm) ref. GT50-03 \footnote{\url{https://www.thorlabs.com/thorproduct.cfm?partnumber=GT50-03}} (hereafter called Blazed300). Both  were placed on the CTIO $0.9\,$m telescope filter-wheel for extensive quasi-simultaneous tests and comparisons.
The images were focused on a $2048\times 2048$ CCD, with pixel size of $24\,\mu$m ($0.4$ arcsec on sky).

\begin{figure}
\begin{center}
\includegraphics[width=\columnwidth]{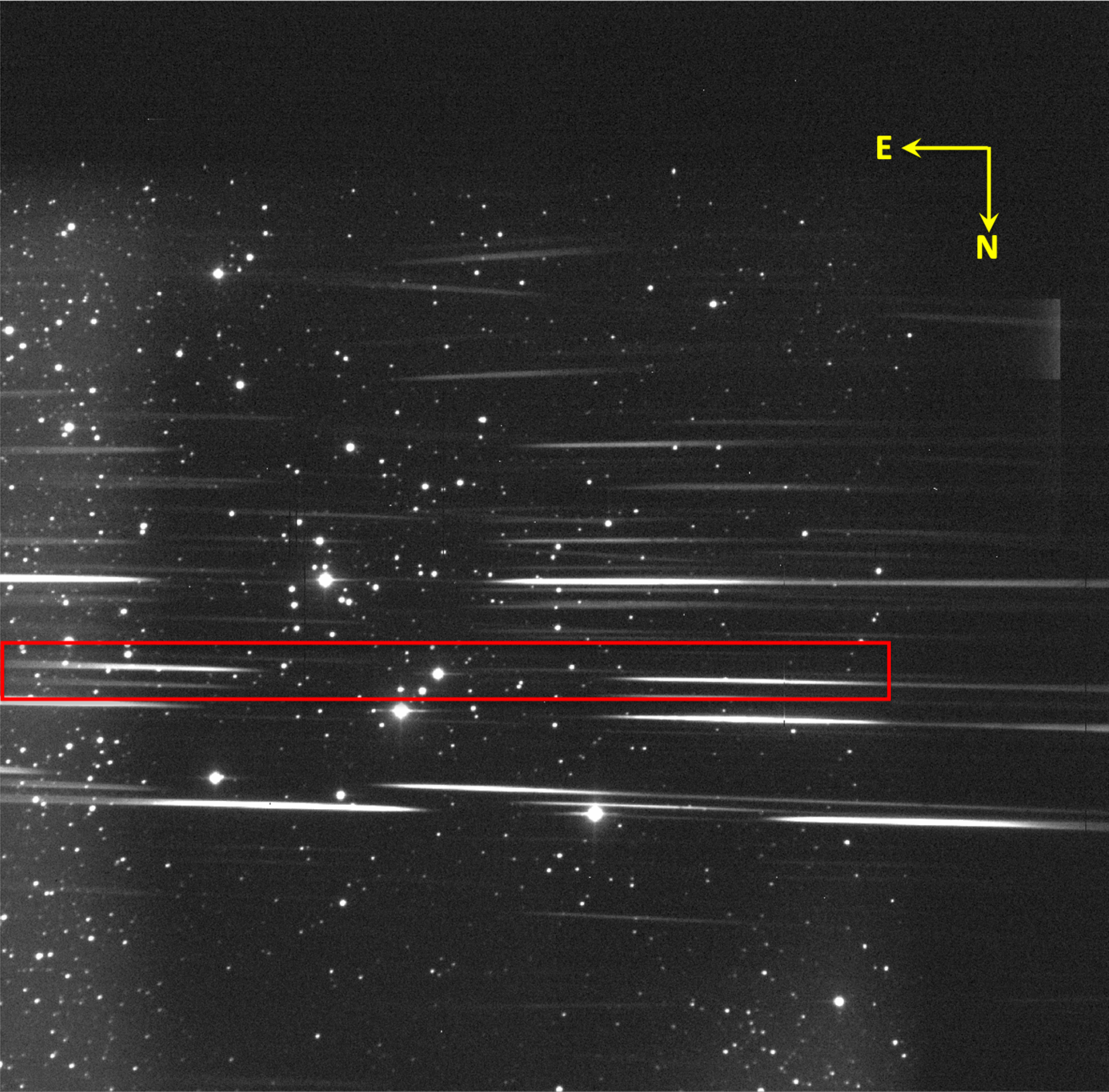}
\includegraphics[width=\columnwidth]{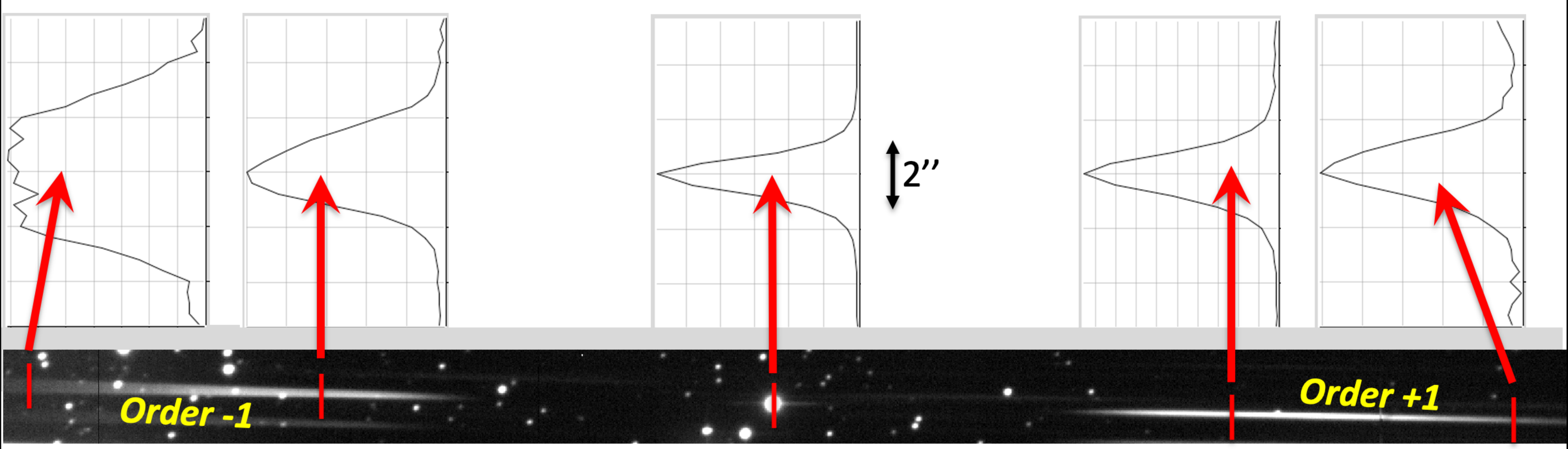}
\end{center}
\caption[] 
{Spectra of stars within the wide open cluster NGC4755, obtained at the CTIO $0.9\,$m telescope with the amplitude hologram. The field is $\sim 13.7'$ wide. $+1$ order spectra are on the right and $-1$ order spectra on the left.
The region in the red rectangle is zoomed in the bottom panel, also showing profiles perpendicular to the dispersion axis for the zero-th order image (center) and at wavelengths $\lambda\sim 565$\,nm and $920$\,nm for both $-1$ and $+1$ orders, allowing to compare the focused (order $+1$) versus defocused (order $-1$) situations.
}
\label{open-cluster}
\end{figure}

We rapidly visually checked some of the specific properties of the holograms by observing the open cluster NGC4755 (see Fig.~\ref{open-cluster}): for each star, we observe a $+1$ order spectrum (on their right) and a $-1$ order spectrum (on their left). The orientation of the dispersion axis depends on the position of the star with respect to the hologram optical center as expected from Fig.~\ref{fig:Neff_alpha} and Eq.~\ref{eq:alpha}. Moreover one can observe from the expanded region that for the $+1$ order spectrum, the focus looks correct from the blue edge (left) to the red edge (right) of the spectrogram, while the $-1$ order  spectrum is defocused, as expected.

\subsection{Measurement of $D_{\mathrm{CCD}}$}

The first step is to check the geometry of the CTIO $0.9\,$m telescope. To do so, we performed $(RA, DEC)$ scans of the Ronchi400 and Blazed300 gratings using an isolated star and a $H_\alpha$ filter in a second filter wheel whose central wavelength is around $\lambda_{H_\alpha} = 656\,$nm. In the images we fit the centroids of the orders 0 and $+1$ using Gaussian profiles and compute the distance $d(\lambda=\lambda_{H_\alpha})$ between both centroids $S_0$ and $S_1(\lambda)$. Assuming the groove density were exactly those given by the makers with an uncertainty of $\pm 1\,$lines/mm, we invert equation~\ref{eq:grating} following notations from Fig.~\ref{fig:dispersion} and Fig.~\ref{fig:frame}:
\begin{align}
D_{\mathrm{CCD}} & = \frac{u_1(\lambda) }{\tan \theta_1(\lambda)} = \frac{u_1(\lambda) }{\tan \arcsin (N_{\rm eff} \lambda+\sin\theta_0)} \notag \\ 
& =u_1(\lambda) \sqrt{(N_{\rm eff} \lambda+\sin\theta_0)^{-2}-1}.
\label{eq:dispersion}
\end{align}
Neglecting angle $\theta_0$ (of the order of arcmin)
compared with $\theta(\lambda)$ (larger than $10\degree$),
we find $D_{\mathrm{CCD}} = 55.45 \pm 0.19\,$mm when averaging the results from the Ronchi400 and Blazed300 gratings. This value is slightly smaller than the $58\,$mm assumed and required for the hologram production.
The thickness variations of the home-made frames installed on the filter wheel also induced variations of the true grating plane position along the telescope axis. We therefore expect fluctuation of this value to be of order $1$ or $2\,$mm between dispersers.
The  $D_{\mathrm{CCD}} = 55.45\,$mm value 
will be used only as a first guess to estimate the dispersion properties of the holograms in the next section; but when astrophysics spectra will be extracted and calibrated in the following of the paper, $D_{\mathrm{CCD}}$ will be fitted from the positions of known emission or absorption lines.

\subsection{Hologram symmetry center and orientation}

\begin{figure}
\begin{center}
\includegraphics[width=\columnwidth]{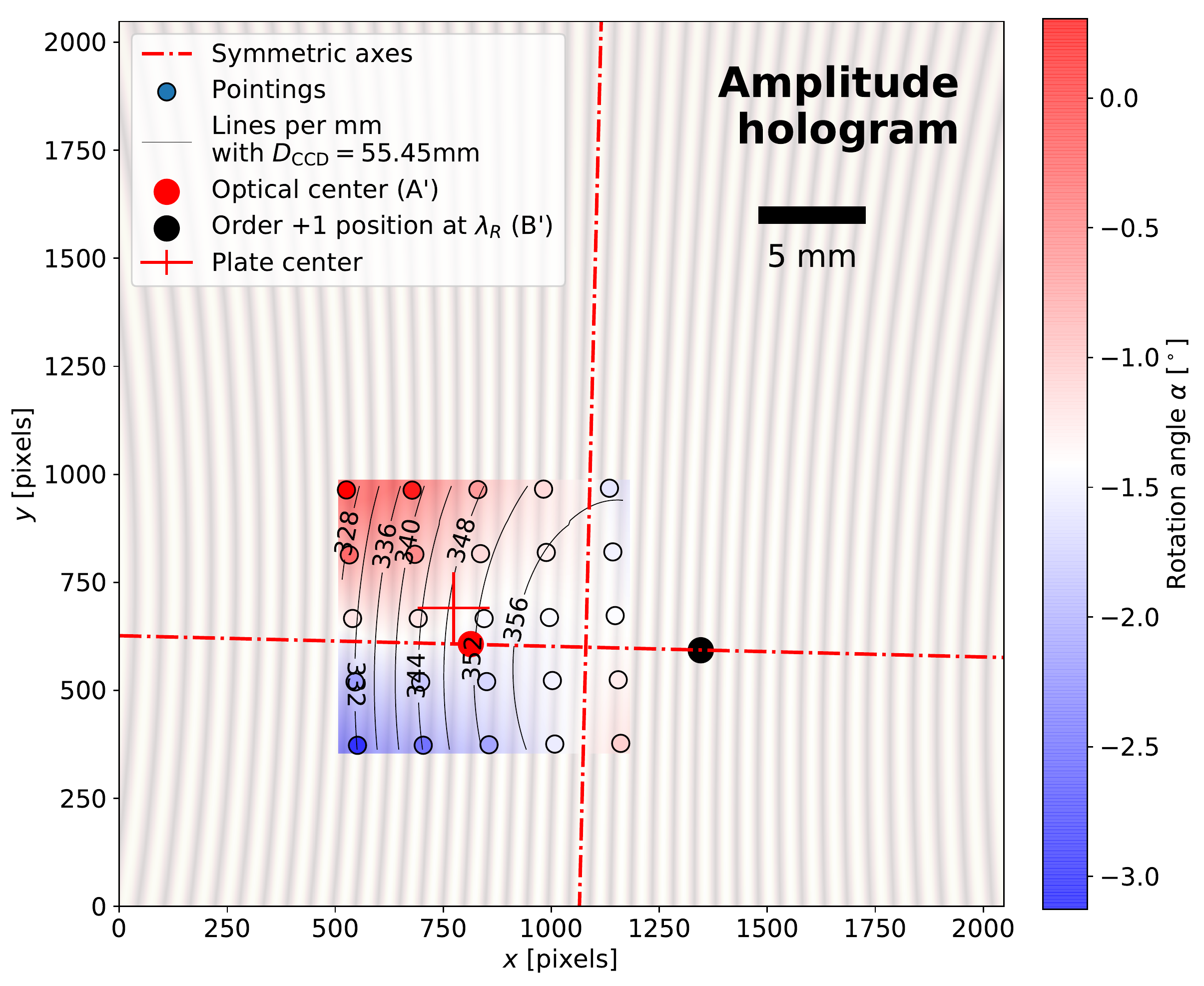}
\end{center}
\caption[] 
{Measurements of the geometrical characteristics of the amplitude hologram. The small colored inset (projected on the CCD) maps the dispersion axis angle $\alpha$ of the $+1$ diffraction spectrum line with respect to the $x$-axis. The open circles correspond to the impacts of the zeroth-order. The fitted saddle point coincides within $2\,$mm with the symmetric center of the holographic pattern (at the crossing of the red dotted lines, as in Fig.~\ref{prod-holo});
the expected optical axis from the makers' indications (red cross)
is supposed to be confused with the big red dot.
Iso-density lines of $N_{\rm eff}(x_0,y_0)$ are also superimposed (in lines/mm) (see Sect.~\ref{sec:dispersion}).
The brown background lines reproduce the holographic pattern of Fig.~\ref{pattern}. }
\label{geometrie-holo}
\end{figure}

The dispersion axis angle $\alpha$ of a spectrum image with respect to the CCD orientation increases  with the vertical distance of the star to the horizontal symmetry axis (see Fig.~\ref{fig:Neff_alpha} and Fig.~\ref{open-cluster}). Therefore, we decided to exploit these inclination angles to determine the true position of the optical center in the CCD pixel coordinates.
To do so, we performed a systematic $(RA, DEC)$ scan of the hologram, with the telescope pointing an isolated star. Given the 0-th order coordinates $S_0(x_0, y_0)$, the angle $\alpha(x_0,y_0)$ of the dispersion axis with respect to the horizontal axis of the CCD is estimated using a Hessian analysis inspired by the interstellar filament detection algorithm used by the \textit{Planck} collaboration in \citet{PlanckFilament}. The elongated structures of the spectrum in the CCD image are detected, and their orientation $\alpha$ are analytically computed (see Appendix \ref{appendix-hessian} for details).
The hologram symmetry center $C$ on the CCD is then given by the saddle point of the $\alpha(x_0, y_0)$ map (see the small colored map on Fig.~\ref{geometrie-holo}). For each hologram, we computed this $\alpha(x_0, y_0)$ map with $5\times 5$ pointings around the expected position of the optical center $A'$, and determined the location of the saddle point $C'$. We were also able to measure the tilt angle $\alpha_0$ of the hologram symmetric axis $(A'B')$ with respect to the CCD horizontal axis $x$. The other symmetric axis is then orthogonal to this axis and passes through the $C'$ point (see the red lines in Fig.~\ref{pattern}).

\subsection{Dispersion properties}\label{sec:dispersion}

The next step to characterize the holograms consists in studying their dispersion properties with the grating formula~\ref{eq:dispersion_fermat}. The incident beam angle
$\theta_0$ is computed using the zero-th order image distance from the CCD center\footnote{At the CTIO $0.9\,$m telescope, the maximum $\theta_0$ value is around 10\,arcmin.}, using a scale of $0.401\,$arcsec/pix so there is a bijective description of the hologram properties in terms of $S_0$ or $S'_0$.

\subsubsection{Measurement of $N_{\mathrm{eff}}$}

$(RA, DEC)$ scans are performed with the silver halide holograms through a $H_\alpha$ filter, similarly to the Ronchi400 and Blazed300 gratings.
After determining the centroids of the orders 0 and $+1$ fitting Gaussian profiles,
using the $D_{\mathrm{CCD}}$ value of 55.45\,mm, 
formula~(\ref{eq:dispersion_fermat}) is inverted for each pointing position $S_0( x_0, y_0,0)$ to find $N_{\mathrm{eff}}(x_0,y_0)$, the local effective number of lines per mm of the hologram, as a function of the undeflected beam impact :
\begin{align}
N_{\rm eff}(x_0,y_0) & = \frac{1}{\lambda_{H_\alpha}} \left[ \sin \left(\arctan  \frac{u_1(\lambda_{H_\alpha} \vert\;  x_0,y_0)}{D_{\mathrm{CCD}}}\right) - \sin\theta_0 \right]
\end{align} 
In the inset of Fig.~\ref{geometrie-holo}, the black contours are the $N_{\rm eff}(x_0,y_0) $ equal-lines for the amplitude hologram.
For both holograms, $N_{\rm eff}$ is around 360 lines/mm at the hologram symmetry center, slowly decreasing toward the edges as expected from Fig.~\ref{fig:Neff_alpha} or Fig.~\ref{pattern}.
Similar results for the phase hologram are presented in Appendix~\ref{appendix_phase}.

\subsubsection{Optical center position}

To find the optical center $A'$, we use the hologram horizontal symmetry axis, the saddle point position $C'$ and the $N_{\rm eff}(x_0,y_0) $ map interpolated at all CCD positions. Given these ingredients, we find the  position along the horizontal symmetry axis where $N_{\mathrm{eff}}$ is such that the order 0 position $S_0$ and the first order position $S_1(\lambda_R)$ at wavelength $\lambda_R$ are symmetrical with respect to the saddle point $C$ projected on the CCD\footnote{In other words, we look for a pointing $S_0(x_0,y_0,0)$ that gives $C'S'_0=-C'S'_1(\lambda_R)$ given  the dispersion relation~(\ref{eq:dispersion_fermat}) and the map $N_{\mathrm{eff}}(x_0,y_0)$. This pointing is the optical center $A'$ when $D_\mathrm{CCD} =D_R$ and the order $+1$ on the sensor is the projection of $B'$.}.
Doing so, we find for the amplitude hologram $d=S_0S_1(\lambda_R)=12.77\,$mm and $d=12.86\,$mm for the phase hologram. This is shorter than the optical bench value ($d_R=13.5\,$mm) but expected because we found at CTIO a shorter $D_{\mathrm{CCD}}$ distance than $D_R$, estimated independently with the Ronchi400 and Blazed300 gratings (all dispersers are inserted in identical frames in the CTIO filter-wheel and their exit face is at the same distance from the CCD).
The distances are indeed related as follows: $d/D_{\mathrm{CCD}} \approx 12.8/55.5 \approx d_R/D_R \approx 13.7/58$.
Given Fig.~\ref{pattern}, the optical center $A'$ is located at a distance $d/2$ toward the left of the saddle point in the hypothesis that $D_\mathrm{CCD} \approx D_R$.
For each hologram, we found the reconstructed optical center to coincide within $2\,$mm with the makers' indications (the cross in  Fig.~\ref{geometrie-holo}). The position $B'$ of the $\lambda_R$ first order is expected at $d/2$ toward the right of the saddle point. The $A'$ and $B'$ positions are sketched by the red and black dots in Fig.~\ref{geometrie-holo}.

\begin{figure}
\begin{center}
\includegraphics[width=0.8\columnwidth]{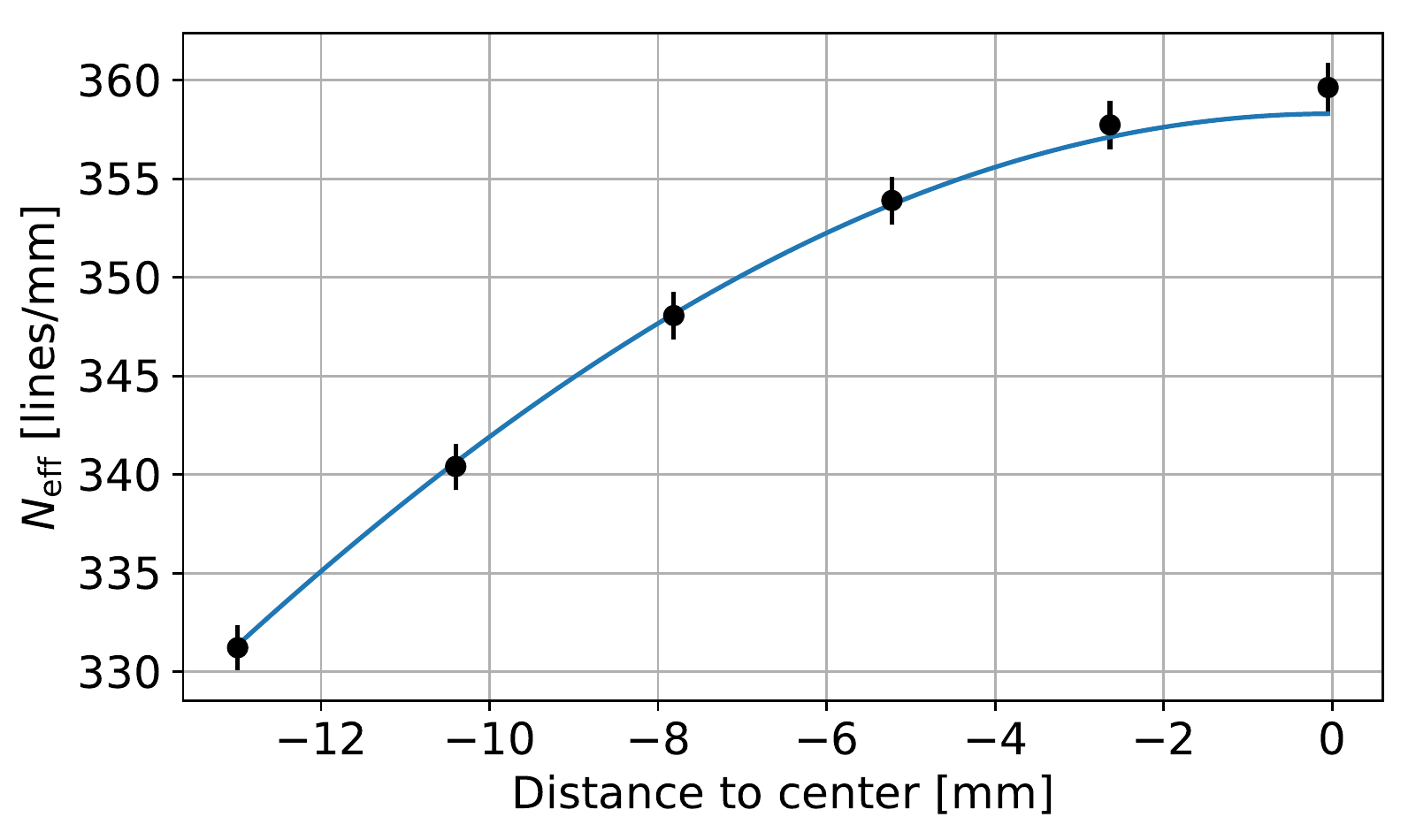}
\end{center}
\caption[] 
{Interpolated values of $N_{\mathrm{eff}}$ along the dispersion axis as a function of the distance to the hologram center (black points) issued from Figure~\ref{geometrie-holo} and the predicted function from equation~\ref{eq:Neff} (blue curve) using the measured $D_{\mathrm{CCD}}=55.45\,$mm and $d=12.8\,$mm distances.}
\label{geometrie-holo-Neff}
\end{figure}

To check the validity of our measurements, we plotted in Fig.~\ref{geometrie-holo-Neff} the interpolated $N_{\mathrm{eff}}$ measurements along the hologram dispersion axis, whose uncertainties are mainly dominated by the $D_{\mathrm{CCD}}$ estimation, and the $N_{\mathrm{eff}}$ predicted curve (formula~\ref{eq:Neff} in Appendix~\ref{app:fermat}) using the measured $D_{\mathrm{CCD}}$ and $d$ distances. We see a very good agreement, convincing us that the holograms we made follow perfectly their expected dispersion performances. 

\subsection{Preliminary extraction of spectra}

Knowing the position of the optical center of the holograms and their dispersion properties, we were able to pursue studies on their performances, still using the CTIO $0.9\,$m telescope, and get calibrated spectra of astrophysical sources. The extraction proceeds as follows:
\begin{enumerate}
    \item estimation of the dispersion axis angle $\alpha$ and rotating the image; since the angles are small, the rotation does not induce significant correlations between the pixel values along the dispersion axis;
    \item fit of the zero-th order centroid; 
    \item crop of the +1 order spectrogram in the image;
    \item fit of a Moffat profile $\phi_{\lambda}(v)$ transverse to the dispersion axis $u$ at all wavelengths $\lambda$ to get the PSF profile and the amplitude of the spectrum; 
    \item wavelength calibration using known absorption or emission bright lines;
    the distance between detected lines
    and tabulated wavelength is minimized to fit again $D_{\mathrm{CCD}}$, with
    given $N_{\rm eff}(x_0, y_0)$, using equation~\ref{eq:dispersion_fermat};
    \item rough flux calibration factor to convert Analog-to-Digital Units (ADU) into erg/s/cm$^2$/nm, accounting for the diameter of the telescope and the CCD gain.

\end{enumerate}
The result of the extraction is a spectrum
\begin{equation}
    S(\lambda) = S_*(\lambda)\times  T_{\rm atm}(\lambda) \times T_{\rm inst} (\lambda),
\end{equation}
where $S_*(\lambda)$ is the astrophysical object spectral energy density (SED), $ T_{\rm atm}(\lambda)$ is the atmospheric transmission, and $T_{\rm inst} (\lambda)$ is the instrumental transmission (including the CCD quantum efficiency). This spectrum corresponds to the first diffraction order, contaminated by the second order light.

\begin{figure}
\begin{center}
\includegraphics[width=\columnwidth]{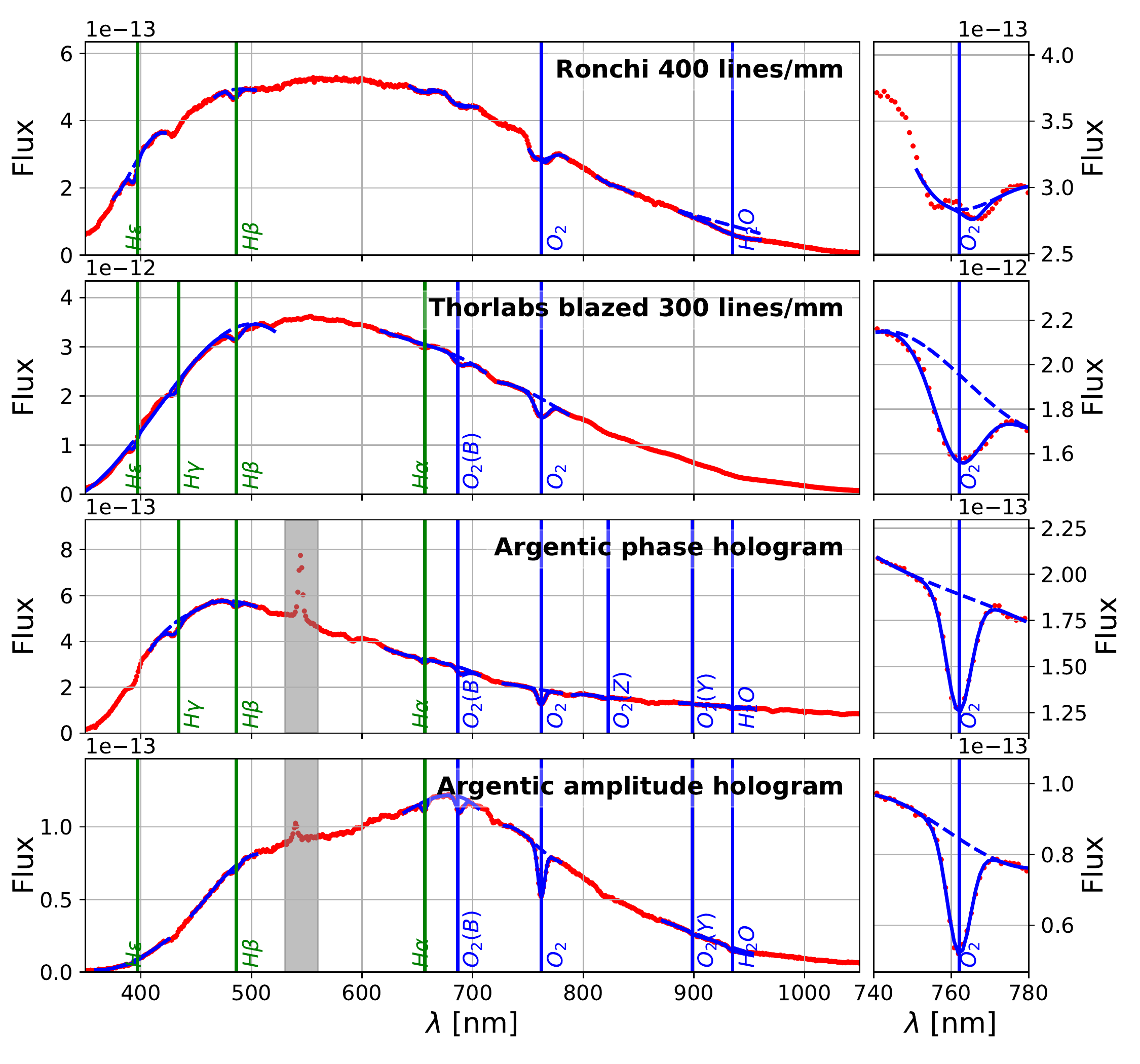}
\end{center}
\caption[] 
{Spectra $S(\lambda)$ in erg/s/cm$^2$/nm units of HD111980 acquired with the four dispersers. Vertical lines show the detected absorption lines with signal-to-noise ratio above 3 and the curved blue lines show the continuum fit (dashed) and the Gaussian fit (plain) to get the lines. The grey zones indicate regions contaminated with field stars (not present in all spectra due to slightly different $\alpha$ angles).  A zoom on the main dioxygen absorption line is provided on the right.}
\label{fig:HD111980}
\end{figure}

In Fig.~\ref{fig:HD111980} we show the spectra of the CALSPEC star HD111980 \citep{Bohlin_2014} observed with identical conditions, extracted with this procedure for the four available dispersers. First, we note that the spectral resolution improves from top to bottom panels (the Ronchi400 grating shows quasi doughnut-shaped absorption lines).
The flux is larger when using the Blazed300 grating and the two holograms show different transmissions and resolutions. All this will be more carefully characterized in the next section.

The extraction of spectra from a slitless spectrograph in a way that is accurate for spectrophotometry is mainly a deconvolution problem that has been addressed in the Spectractor software\footnote{\url{https://github.com/LSSTDESC/Spectractor}} \citep{2021ascl.soft04004N} with forward modelling.
The recipe consists in simulating a raw image from the model of the instrument and the atmosphere that best fits the observed data, rather than manipulating the data image to obtain the spectral information.
For the direct extraction of atmospheric transmission from stars with a known spectrum, the instrumental transmission must also be precisely known. For the spectra presented in this paper (as in  Fig.~\ref{fig:HD111980}), using an accurate 2D PSF function that models the defocusing of the dispersers or even the donut shape of the Ronchi grating PSF can improve the spectral resolution and photometric accuracy for all spectra, but this model must be accurate. The data set to model correctly the PSF shape can be acquired with an optical bench in a laboratory, or looking at sky objects with strong emission lines, but this was not available at the time of the analysis described here. For the proof-of-concept of the holographic dispersers presented in this paper, we used the simple extraction of the spectra described in this section, and we will continue to refine the extraction technique in our next papers.

\section{Performances of the holographic optical elements}
\label{Sec:performances}
\subsection{Focusing performances}
\label{sec:focus}

The first purpose of our observational tests was to check the improvement of the hologram focus for the red and IR light with respect to a periodic grating. 

\begin{figure}
\begin{center}
\includegraphics[width=0.75\columnwidth]{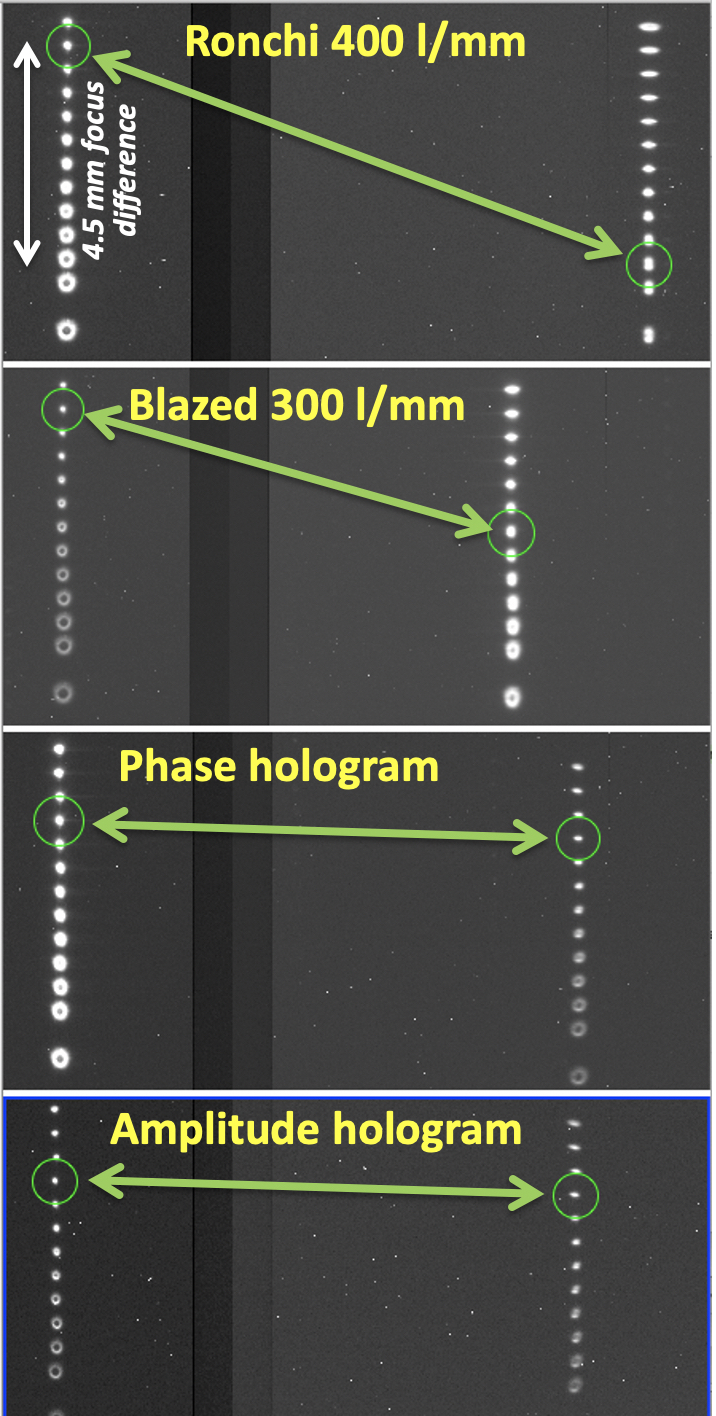}
\end{center}
\caption[] 
{Focus procedures with the $H_{\alpha}$ narrow filter. The series of horizontal pairs of spots show the successive images obtained when changing the telescope focusing (by $0.5$\,mm steps). From top to bottom, Ronchi400, Blazed300, phase and amplitude holograms ($\sim\,$350 lines/mm). The zero order is on the left side, and the corresponding first order is on the right side. The Ronchi400 shows the largest difference between the best focus distance for the two orders in $H_{\alpha}$ ($4.5\pm 0.5$\,mm).}
\label{focus}
\end{figure}

To get a first idea of this improvement, we observed a star through a $H_\alpha$ narrow filter (central wavelength $656\,$nm, FWHM$=6.4\,$nm) and used the standard focus procedure of the telescope to systematically compare the 0 and +1 orders focuses for the gratings and the holograms.
Here, we define the +1 order focus as the position of the waist along the dispersion axis, in order to minimize the wavelength mixing when observing a continuum spectrum.
As shown in Fig.~\ref{focus}, we found the +1 order focus of the holograms at the same position as the zero-th order focus within 0.5\,mm accuracy. In contrast, we found a $4.5\pm 0.5\,$mm difference between the 0 and +1 orders for the Ronchi400, and  about $2.5\pm 0.5\,$mm difference for the Blazed300 (due to its smaller $N_{\rm eff}$ value). These values are compatible with the shifts expected by the simulations (see Fig.~\ref{defocus} for the Ronchi400).

\begin{figure}
\begin{center}
\includegraphics[width=\columnwidth]{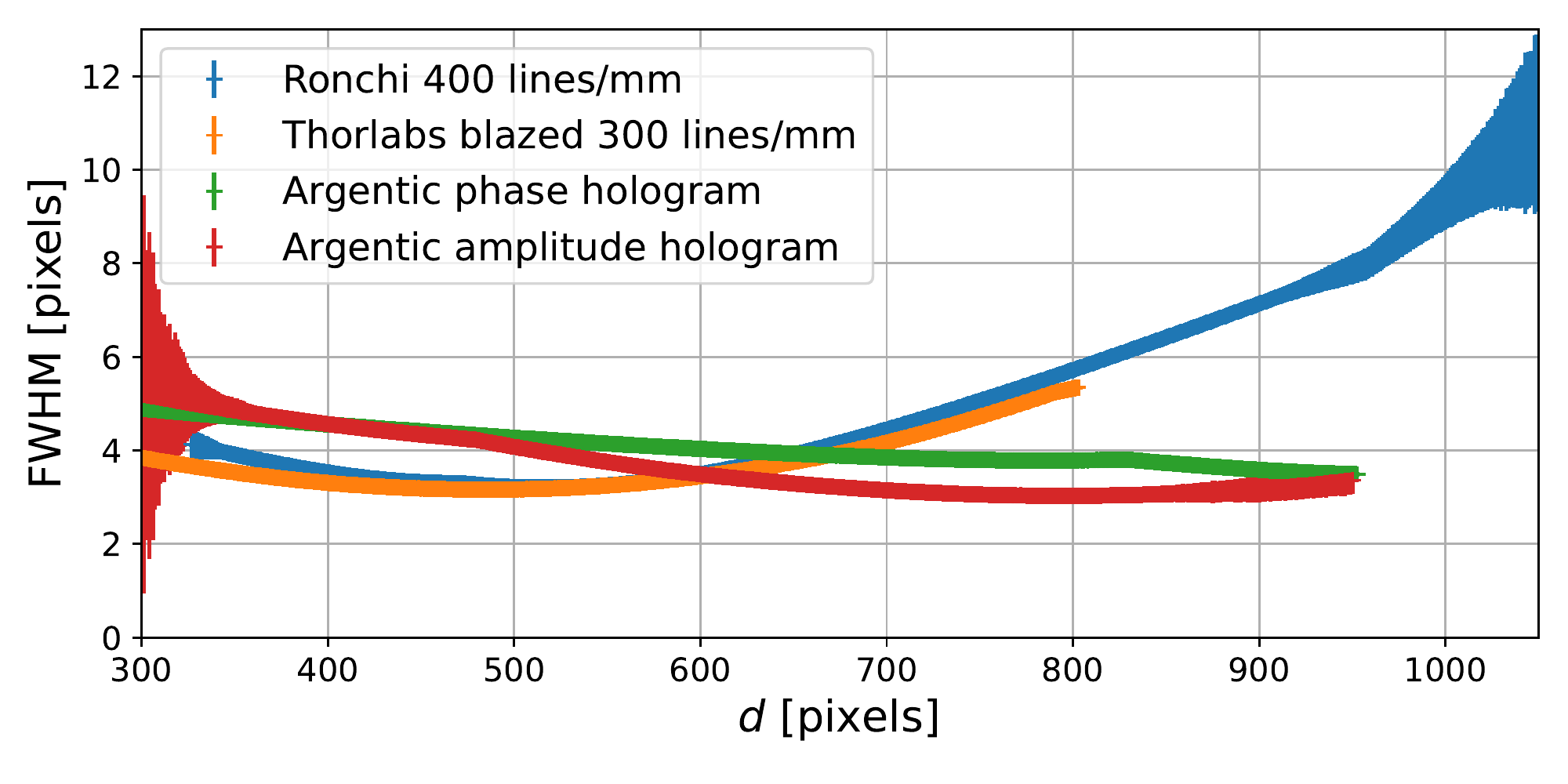} \\
\end{center}
\caption[]{FWHM of the transverse profile of the four spectra from Fig.~\ref{fig:HD111980}, as a function of the distance to the order 0 centroid $d(\lambda)=S_0S_1(\lambda)$.
The seeing provided by the CTIO seeing monitor was $\sim 0.7$ arcsec {\it i.e.} 1.8 pixel.
The width of the bands represents the FWHM uncertainty due to the flux measurement uncertainty.
Note that since dispersions change from one disperser to another, the same $d$ abscissa corresponds to different wavelengths.
}\label{profiles}
\end{figure}

The focusing performances shown in Fig. \ref{focus} are quantitatively confirmed by the analysis of the width of the fitted cross-sectional profile $\phi_{\lambda}(v)$. In Fig. \ref{profiles} is represented the full width half maximum (FWHM) of the spectrum transverse profile as a function of the distance to the order 0 centroid $d(\lambda)=S_0S_1(\lambda)$. This quantity is evaluated directly on the median of a stack of ten consecutive transverse profiles. The seeing was $\sim 0.6$ arcsec {\it i.e.} 1.5 pixel,
with variations from $\lambda=400\,nm$ to $1000\,nm$ not exceeding $0.25$ pixel.
The Ronchi400 and Blazed300 gratings show similar behaviour with an increasing FWHM with $d(\lambda)$ as expected.
The hologram gratings present a rather constant cross-sectional FWHM along the dispersion axis, with a better focus in the red, a consequence of the fact that they were recorded at  $\lambda_R = 639\,$nm.
We conclude that the holographic prototypes have the expected performance concerning their focusing power, in particular in the reddest part where the water vapor absorption band stands.

\begin{figure}
\begin{center}
\includegraphics[width=\columnwidth]{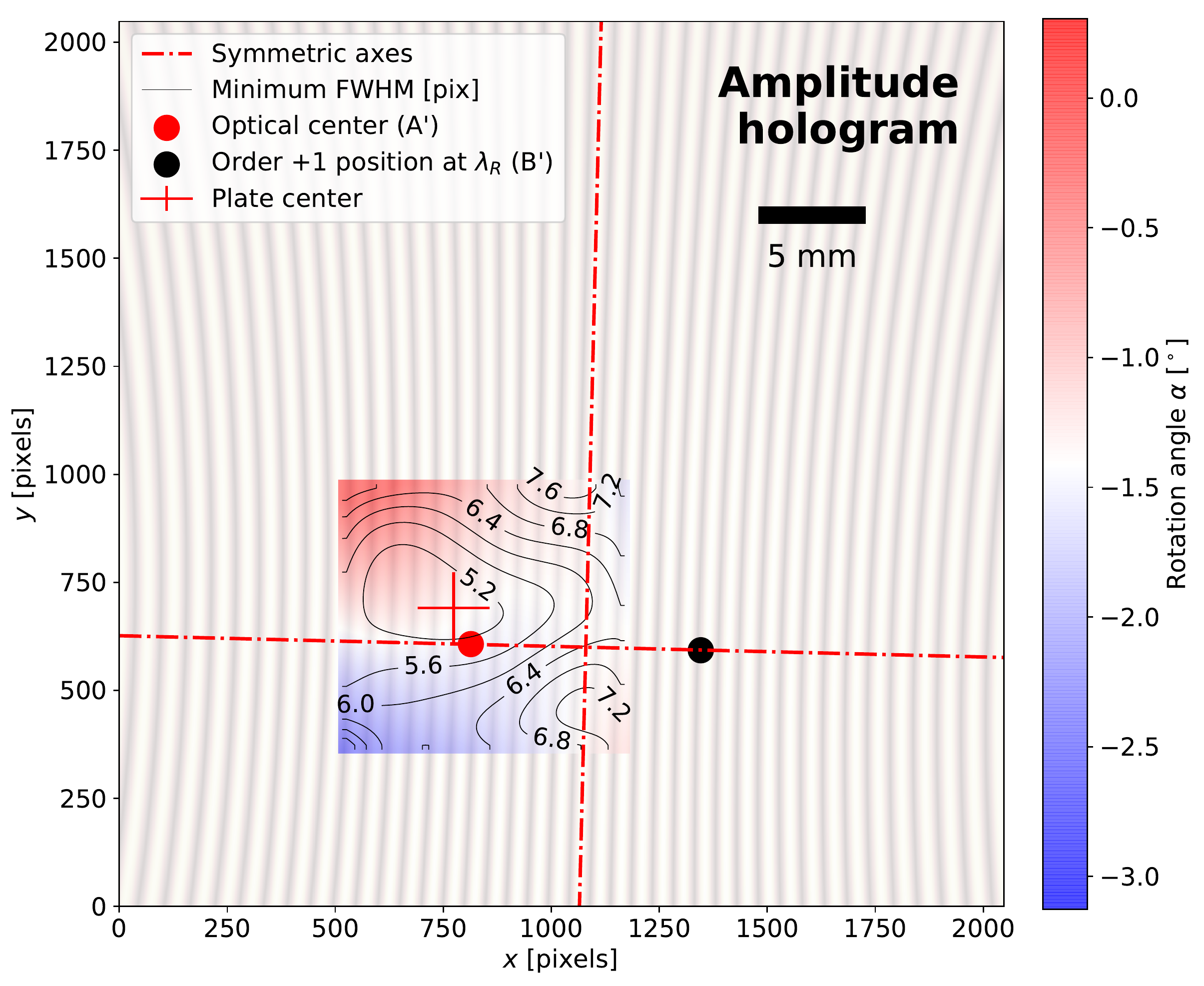}
\end{center}
\caption[] 
{Measurements of the focusing characteristics of the amplitude hologram.
Iso-lines of the minimum FWHM (in pixel) respectively to the wavelength, as a function of the order 0 position,
superimposed on the same frame than Fig.~\ref{geometrie-holo}. The seeing provided by the CTIO seeing monitor was about $1.2\,$arcsec {\it i.e.} around 3 pixels.
}
\label{geometrie-holo-fwhm}
\end{figure}

One important test of the holograms is to check if the PSF is degraded when the order 0 image does not coincide with the optical center as determined in Sect.~\ref{Sect:telescope}. We performed again $5\times 5$ scans of telescope pointings while observing CALSPEC stars without any filter, and extracted the spectra.
All these measurements were performed together during a short time interval (1h15), thus limiting the variations of the atmospheric conditions.
In Fig.~\ref{geometrie-holo-fwhm}, we report the minimum FWHM observed along the spectra as a function of the stellar order 0 image position.
We observe that this minimum FWHM degrades by less than 1 pixel ($24\,\mu$m or $0.401\,$arcsec) when the order 0 position is moved by less than $200$\,pixels from the optimal position on the CCD. This pixel domain is contained in a box of about 1\,cm$\times$1\,cm, large enough to allow for a comfortable use of these holograms and maintain high focusing performances.

\subsection{Resolution performances}
The spectral resolution is defined as
$R(\lambda) = \lambda/\Delta\lambda_{min}$ where $\Delta\lambda_{min}$ is the minimal detectable separation between monochromatic lines.
\subsubsection{Theoretical best resolution of a slitless spectrograph}
Two spectroscopic lines from a source can be separated when their dispersion on the sensor exceeds the extension of the spot of one monochromatic line.
Consider a perfectly monochromatic source which produces a direct image of angular size $\sigma_0$ on the detector. The disperser produces a first-order diffracted image with a spread of $\sigma_1$ along the dispersion direction.
We want to know, for this same source, what is the chromatic width $\Delta\lambda_{min}$ which produces a first-order diffracted image with the same spread $\sigma_1$.
We then assume that $\Delta\lambda_{min}$ can be equated to the power of separation, below which the diffracted image does not differ from an image of monochromatic source.
Starting from the initial causes of zero-th order image spread ($\sigma_0$), we have to estimate the first-order diffracted image spread ($\sigma_1$) to deduce $\Delta\lambda_{min}$.

Since our spectrograph is slitless, there is no slit width to consider, and $\sigma_0$ depends on the seeing, the focus quality, the distortions, the pixel size and the angular size of the source.
In the optimal cases (point-source, pixel size $\ll$ seeing, perfect focus, no distorsion), $\sigma_0$ is dominated by the atmospheric seeing.
Eventually the variation of the seeing with the wavelength can be taken into account, since it is scaling as $\lambda^{-1/5}$ \citep{Boyd_78}, decreasing by a factor 0.83 from $400\,$nm to $1000\,$nm.

Following notations of Fig.~\ref{fig:dispersion},
when the direction of an incoming ray $\theta_0$ changes by $\sigma_0$ (due to one of the causes just mentionned), then the impact of the telescope beam on the grating $S'_0$
varies by $\sigma_0\times (f-D_{\mathrm{CCD}})/\cos^2{\theta_0}$, where $f$ is the focal length of the telescope.
The projection on the sensor plane $S_{\perp}$ also varies by the same quantity.
Taking the derivative of Eq.~(\ref{eq:dispersion}) with respect to $\theta_0$ and using the derivative of Eq.~(\ref{eq:grating}), one finds
$\delta(S_{\perp}S_1) = \sigma_0\times D_{\mathrm{CCD}}\cos{\theta_0}/\cos^3{\theta_1(\lambda)}$.
Then the shift of $S_1$ is:
\begin{equation}
    \delta S_1(\lambda) = \sigma_1.f = \sigma_0\times \left( \frac{f-D_{\mathrm{CCD}}}{\cos^2{\theta_0}} + \frac{D_{\mathrm{CCD}}\cos{\theta_0}}{\cos^3{\theta_1(\lambda)}}\right).
\end{equation}
In our case (and in general), $D_{\mathrm{CCD}} \ll f$ and $\theta_0 <$ few arcmin, and the consequence is that $\sigma_1\sim \sigma_0$.
Incidentally, this is also true in the direction perpendicular to the dispersion axis.
It should be noted that this result concerns the chief ray of the telescope beam (the axis of the light cone), and is therefore only valid if the first order diffracted image is focused like the direct image.

Now we examine the spreading of the image in $S_1$ under the sole effect of a chromatic broadening of the source.
When shifting by $\Delta\lambda$ the wavelength of a point source,
only the impact $S_1(\lambda)$ of the first order diffracted beam is shifted
($\theta_0$ and consequently $S'_0$ and $S_{\perp}$ are unaffected).
We obtain the shift $\delta S_1(\lambda)$
by derivating Eq.~(\ref{eq:dispersion}) with respect to $\lambda$, and using Eq.~(\ref{eq:grating}) to get the derivative of $\theta_1(\lambda)$ with respect to $\lambda$.
Finally, the shift of $S_1$ is:
\begin{equation}
    \delta S_1(\lambda) = \Delta\lambda \times \frac{N_\mathrm{eff} D_{\mathrm{CCD}}}{\cos^3{\theta_1(\lambda)}}.
\end{equation}
Our hypothesis states that if $\delta S_1(\lambda) = \sigma_1.f (= \sigma_0.f)$, then $\Delta\lambda = \Delta\lambda_{min}$
and we get the relation
\begin{equation}
\Delta\lambda_{min}(\sigma_0)=f\sigma_0\cos^3{\theta_1(\lambda)}/(N_\mathrm{eff} D_{\mathrm{CCD}}).
\end{equation}
The wavelength separation between two spectral lines has to be larger than $\Delta\lambda_{min}(\sigma_0)$ to avoid the confusion due to the direct image spread characterized by $\sigma_0$.
Assuming normal incidence ($\sin{\theta_0} = 0$) and expressing $\cos{\theta_1(\lambda)}$ from Eq.~(\ref{eq:grating}),
the theoretical resolution of our slitless spectrometer with imaging quality limited by the dispersion $\sigma_0$
is finally :
\begin{equation}
    R(\lambda,\sigma_0) = \frac{\lambda}{\Delta\lambda_{min}(\sigma_0)} = \frac{D_{\mathrm{CCD}}}{f \sigma_0} \frac{\lambda N_\mathrm{eff}}{[1-(\lambda N_\mathrm{eff})^2]^{3/2}}.
    \label{eq:resoltheor}
\end{equation}

\subsubsection{Measured resolution performances}
For the CTIO $0.9\,$m telescope with $f=21.6\,$m and for holograms with $N_\mathrm{eff}\sim 350\,$mm$^{-1}$ and $D_{\rm CCD}=55.45\,$mm,
the best theoretical resolution is:
\begin{equation}
    R_{theor.}(\lambda,\sigma_0) \sim 337 \left[\frac{\sigma_0}{1\, {\rm arcsec}}\right]^{-1}\left[\frac{\lambda}{1\,\mu \text{m}}\right]\left[1-0.123\left[\frac{\lambda}{1\,\mu \text{m}}\right]^2\right]^{-3/2}.
    \label{eq:resolution}
\end{equation}

\begin{figure}
\begin{center}
\includegraphics[width=\columnwidth]{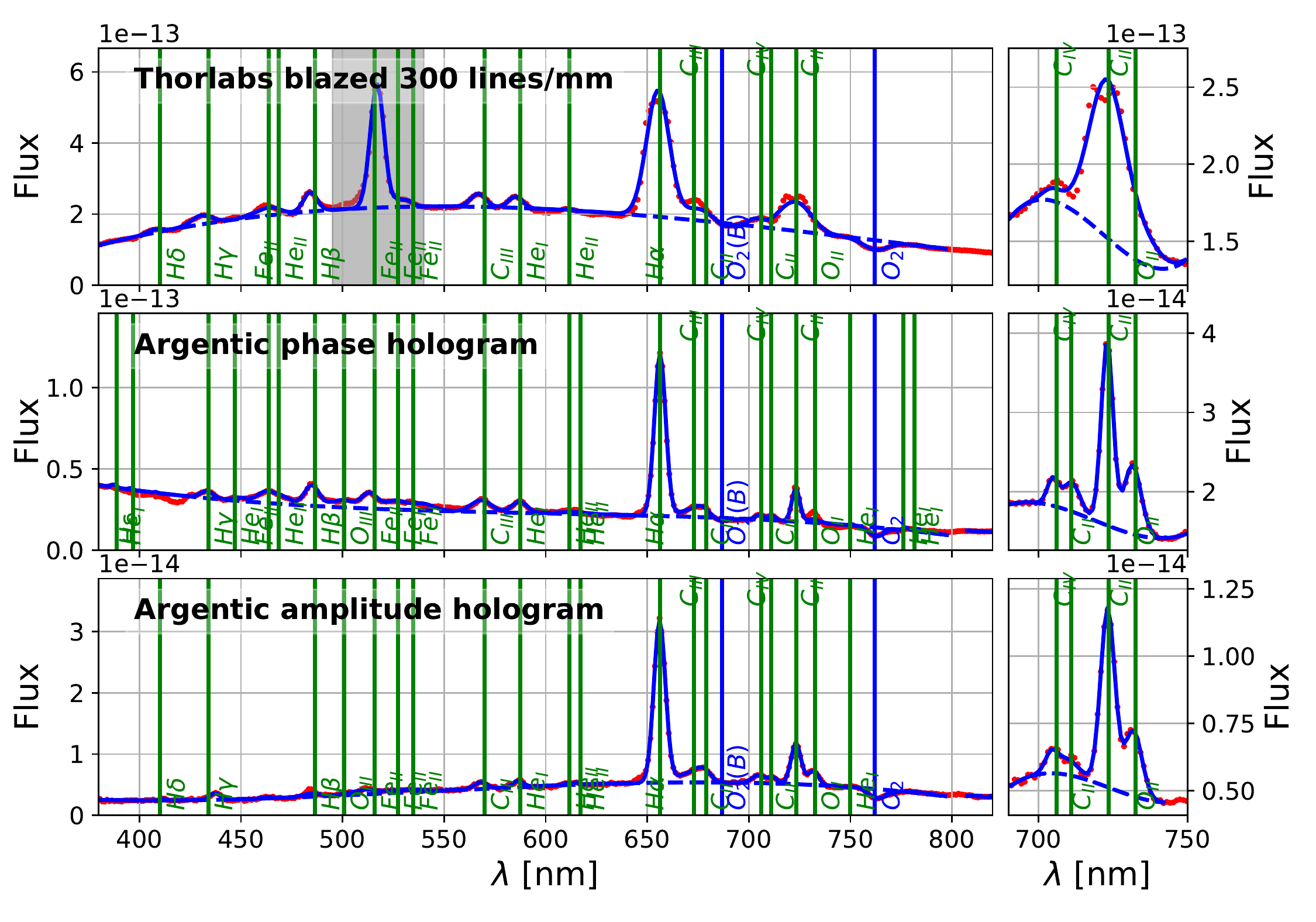}
\end{center}
\caption[] 
{Spectra $S(\lambda)$ from planetary nebula PNG321.0+3.9 (extension $2\,$arcsec) in erg/s/cm$^2$/nm units measured with the Blazed300 grating, the phase hologram, and the amplitude hologram. Vertical green lines show the detected emission lines with signal-to-noise ratio above 3 and the curved blue lines show the continuum fit (dashed) and the Gaussian fit (plain) to the lines. The grey zones indicate regions contaminated with field stars (not present in all spectra due to slightly different $\alpha$ angles). A zoom on CII and CIV carbon emission lines is provided on the right.}\label{spectre-PN}
\end{figure}

We measured the effective spectral resolution of the gratings thanks to the observation of a small angular-size planetary nebula with sharp emission lines. Spectra of the planetary nebula PNG321.0+3.9 (or HEN 2-113), with an angular extension of $\sim 2"$ FWHM \citep{Lagadec_2005}, are presented in Fig.~\ref{spectre-PN}, and the resolution $\Delta \lambda_{min}$ is estimated from the measured width of the $H_\alpha$
\footnote{The $H_\alpha$ line is not contaminated by a $N_{II}$ contribution (wavelengths differing by only $2.1\,$nm) according to \citep{vizier:V/84}; we are able to confirm this fact considering the sharpness of the line obtained with the holograms.}
and $H_\beta$ emission lines (see Table~\ref{tab:resolution}).
Rough estimates of the effective spectrograph resolution $\lambda /\Delta\lambda_{min}(\sigma_0)$ are also quoted in Table~\ref{tab:resolution} for the two spectral lines.

To get indicative information on the performances for redder color, we add in Table~\ref{tab:resolution} the measured width of the atmospheric $O_2$ absorption band at $762\,$nm, as observed in the spectrum of HD111980
(Fig.~\ref{fig:HD111980}). Since this absorption band is wide, the measured $\Delta\lambda$ is wider than $\Delta\lambda_{min}(\sigma_0)$, and it can not be used to estimate the resolution.
The Ronchi400 spectral resolution has not been evaluated as the Gaussian profile fit fails for lines that are too strongly defocused.

We notice that in the red part ($\lambda=659\,$nm), the observed resolutions for the holograms are closer to the estimates of Eq. (\ref{eq:resolution}) (assuming $\sigma_0=2"/2.35$) than for the Blazed300 grating.
Indeed, as the Blazed300 grating doesn't focus correctly at order 1 when the order 0 is focused, then $\sigma_1>\sigma_0$ and the true resolution is worse than expression (\ref{eq:resolution}).

In the bluer part of the spectra, the three gratings are almost equivalent, with performances close to the theoretical expectations, limited here by the nebula size. We observe a slight advantage to the Blazed300 grating and a disadvantage for the amplitude hologram (probably due to a low signal-to-noise ratio in this blue part). However, in the redder part of the spectra, the holograms clearly benefit from their better focus.

\begin{table}
    \begin{center}
\begin{tabular}{ccccc} \hline \hline \\ [-1ex]
     & &            & Phase & Amplitude   \\
     & & Blazed300 & Hologram & Hologram   \\
  $\lambda$  & $N_{\rm eff}$ & $300\,$l/mm & $\sim350\,$l/mm & $\sim350\,$l/mm \\  [1ex] \hline \\ 
            & $\Delta \lambda_{min}$ & $3.4\,$nm  & $3.5\,$nm &  $4.1\,$nm \\ 
$H_\beta\ (483\,\text{nm})$   & $R_{meas.}$   & 145   & 140   & 115 \\
            & $R_{theor.}$  &   160    &   190    &  190 \\  [1ex] \hline \\ 
            & $\Delta \lambda_{min}$ & $5.6\,$nm  & $2.8\,$nm &  $2.9\,$nm \\ 
$H_\alpha\ (659\,\text{nm})$   & $R_{meas.}$   & 115   & 240   & 230 \\
            & $R_{theor.}$  &   230    &   270    & 270 \\  [1ex] \hline \\ 
$O_2\ (762\,\text{nm})$ & $\Delta \lambda$ & $5.9\,$nm  & $3.2\,$nm &  $3.0\,$nm \\ [1ex] \hline \\ 
\end{tabular}
\centering
\caption[]{Resolution of the gratings.
  $\Delta \lambda_{min}$ values are the RMS of the Gaussian profiles fitted to the emission lines ($H_\beta$ and $H_\alpha$); $\Delta \lambda$ is the RMS of the profile of the (wider) $O_2$ absorption band.
The measured spectrograph resolution $R_{meas.}=\lambda/ \Delta\lambda_{min}$ can be compared with the theoretical one $R_{theor.}$, computed from Eq. \ref{eq:resoltheor} for a source of 2 arcsec extension.}
\label{tab:resolution}
\end{center}
\end{table}

\subsection{Sensitivity to atmospheric parameters}
\begin{figure}
\begin{center}
\includegraphics[width=\columnwidth]{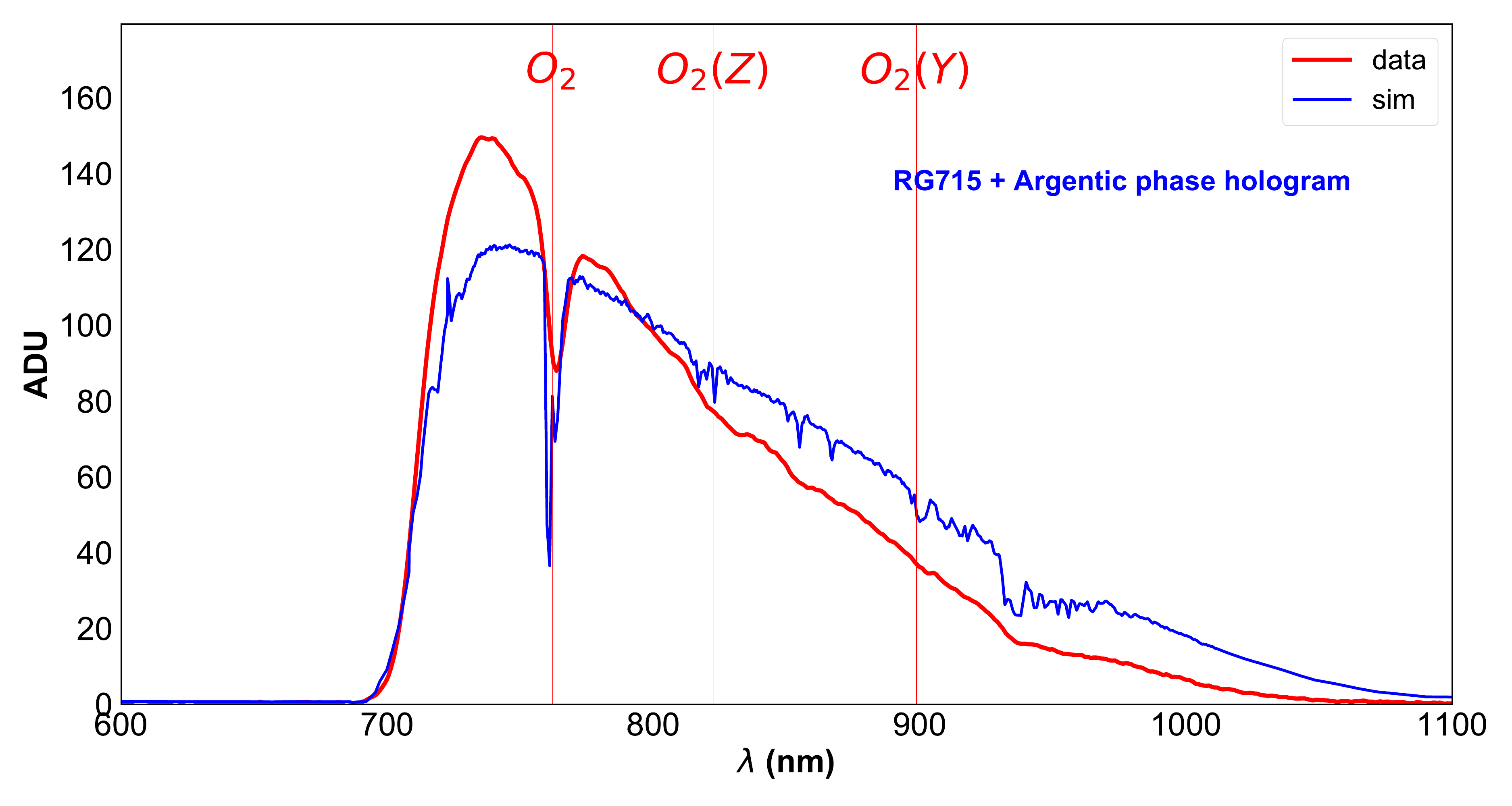}
\begin{tabular}{cc}
\includegraphics[width=0.47\columnwidth]{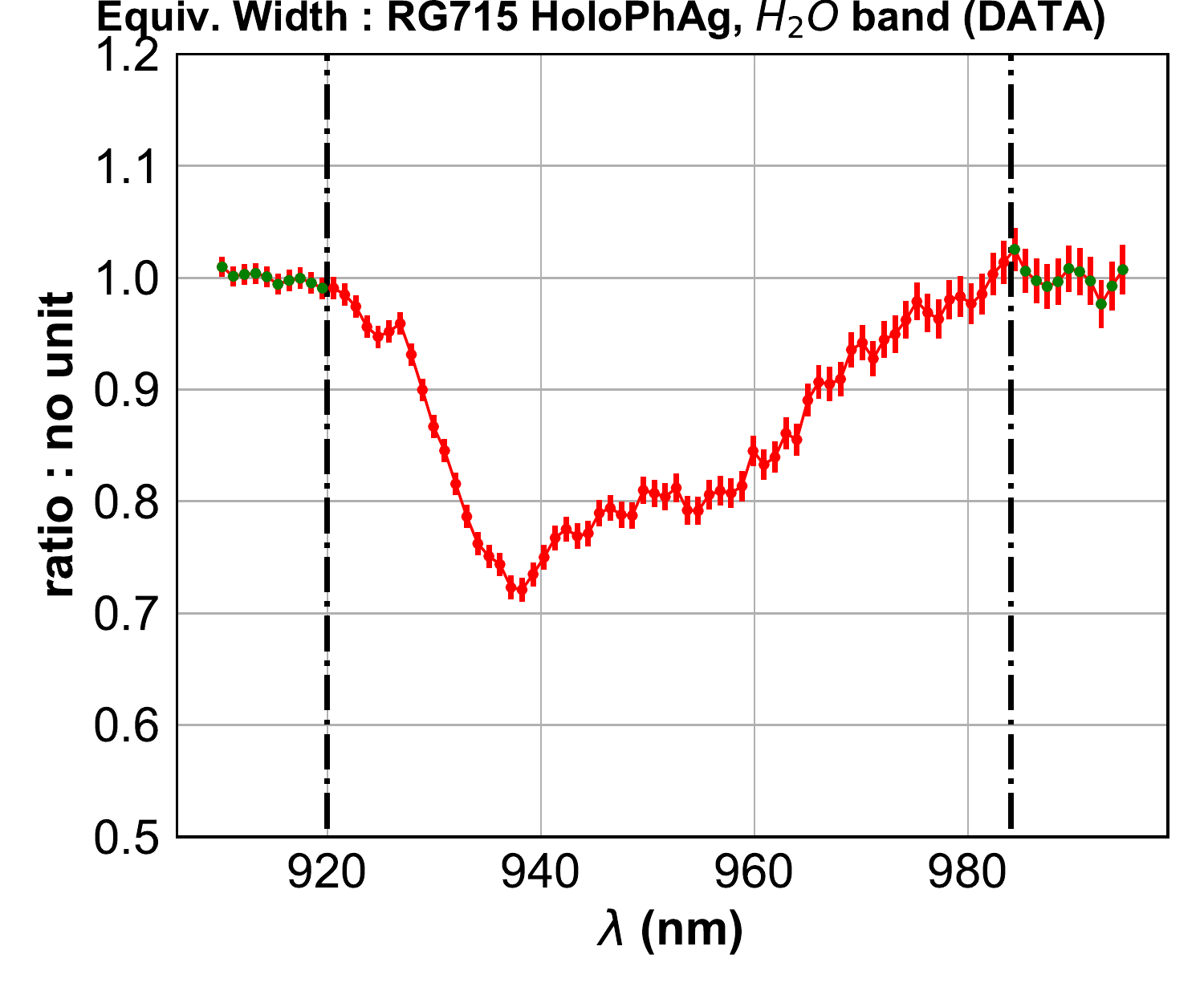} &
\includegraphics[width=0.47\columnwidth]{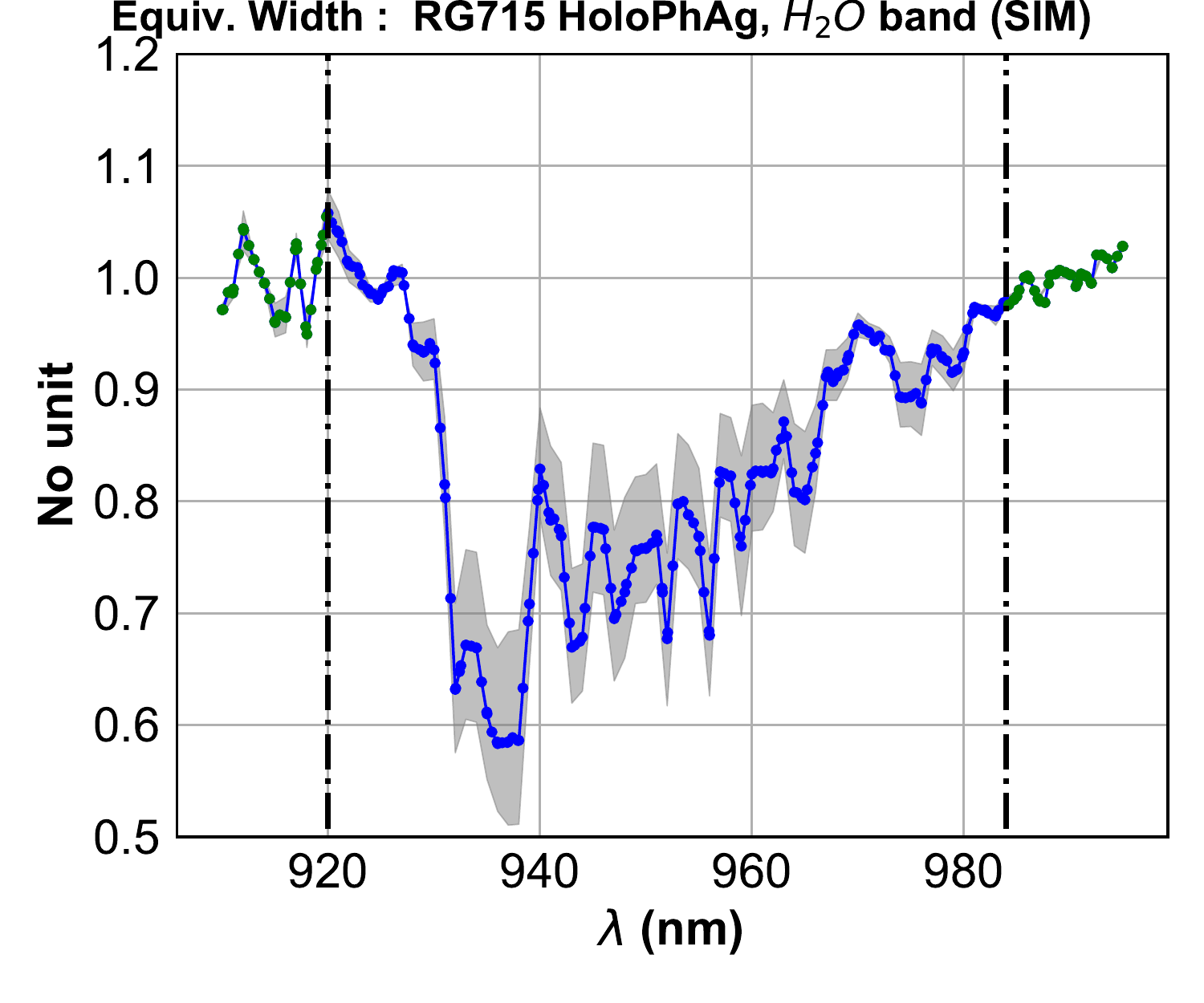}
\end{tabular}

\end{center}
\caption[] 
{The top panel shows the CALSPEC~HD111980 standard spectrum filtered through the low passband filter RG715, obtained with the phase hologram (data in red, simulation in blue, with a rough throughput model). The bottom panels focus on the water vapor absorption band (left is data, right is simulation), showing ratios of the spectrum to the estimated continuum. Blue curve refers to a simulation with precipitable water set at 4~mm. Grey region in simulation refers to a range of precipitable water wapor between 2-6~mm 
(equivalent widths are respectively 7.7, 10.5, 12.7 nm).
}

\label{water}
\end{figure}

Fig. \ref{water} shows the red section of a reduced spectrum of the CALSPEC standard HD111980, with a zoom around the water vapor absorption band ($850-1000\,$nm), one of the main features we plan to use for the Legacy Survey of Space and Time (LSST) atmospheric calibration.
We used here a red filter (RG715) that blocks wavelengths below 700~nm, to avoid superimposition of the second diffraction order blue light with the first order red light.
The equivalent width estimated for the airmass $z=1.26$ of this observation is $EQW_{H_2O}(data)=8.5 \pm 1.5$~nm.
As this equivalent width is not very sensitive to the details of the transmission, as long as it does not vary too abruptly, we can confront it with a simulation.
In the figure, the simulated spectrum has been obtained by multiplying the CALSPEC~HD111980 SED by the atmospheric transmission and by a guessed (rough) typical telescope transmission (optical throughput and CCD quantum efficiency). The atmospheric transmission profile is calculated
by using the Atmospheric Radiation Transfer package LibRadTran \citep{LibRadTran_2015}, setting the typically expected precipitable water vapor at $4\pm 2\,$mm for a ground altitude of $2200\,$m and airmass $z=1.26$.
Although the CTIO $0.9\,$m telescope CCD quantum efficiency is very low in this wavelength domain, we are however able to unambiguously detect the water absorption band, observe that its shape is in good agreement with the expectations, and measure its equivalent width that allows us to roughly estimate the precipitable water quantity (between 2 and 4~mm).
These preliminary results are very promising, since the AuxTel camera will benefit from the same type of CCD than the Rubin Observatory SST Camera, with a much better IR quantum efficiency.

\subsection{Transmission efficiency in the first diffraction order}

The resolution of the spectrum appears significantly better with the holograms, but the Blazed300 grating transmits significantly more light in the first order. Moreover, as can be seen in Fig.~\ref{fig:HD111980}, the two holograms do not have the same transmission functions of the wavelength. As expected, the amplitude hologram, which modulates the light absorption, has a significantly lower transmission than the modulation index hologram, which is essentially translucent.
We found the following rough transmission ratios for the first diffraction order at $760\,$nm:
\begin{itemize}
    \item Blazed Thorlabs300/Phase hologram: $\sim 10$
    \item Phase hologram/Amplitude hologram: $\sim 2$
\end{itemize}
While the transmission is not critical for the purpose of the present paper, the final holograms to be used for a slitless spectrophotometer will be optimized for the best transmission efficiency and uniformity. This point will be specifically addressed in our next paper \citep{Holospec2_2021}.

\subsection{Second diffraction order}

\begin{figure}
\begin{center}
\includegraphics[width=\columnwidth]{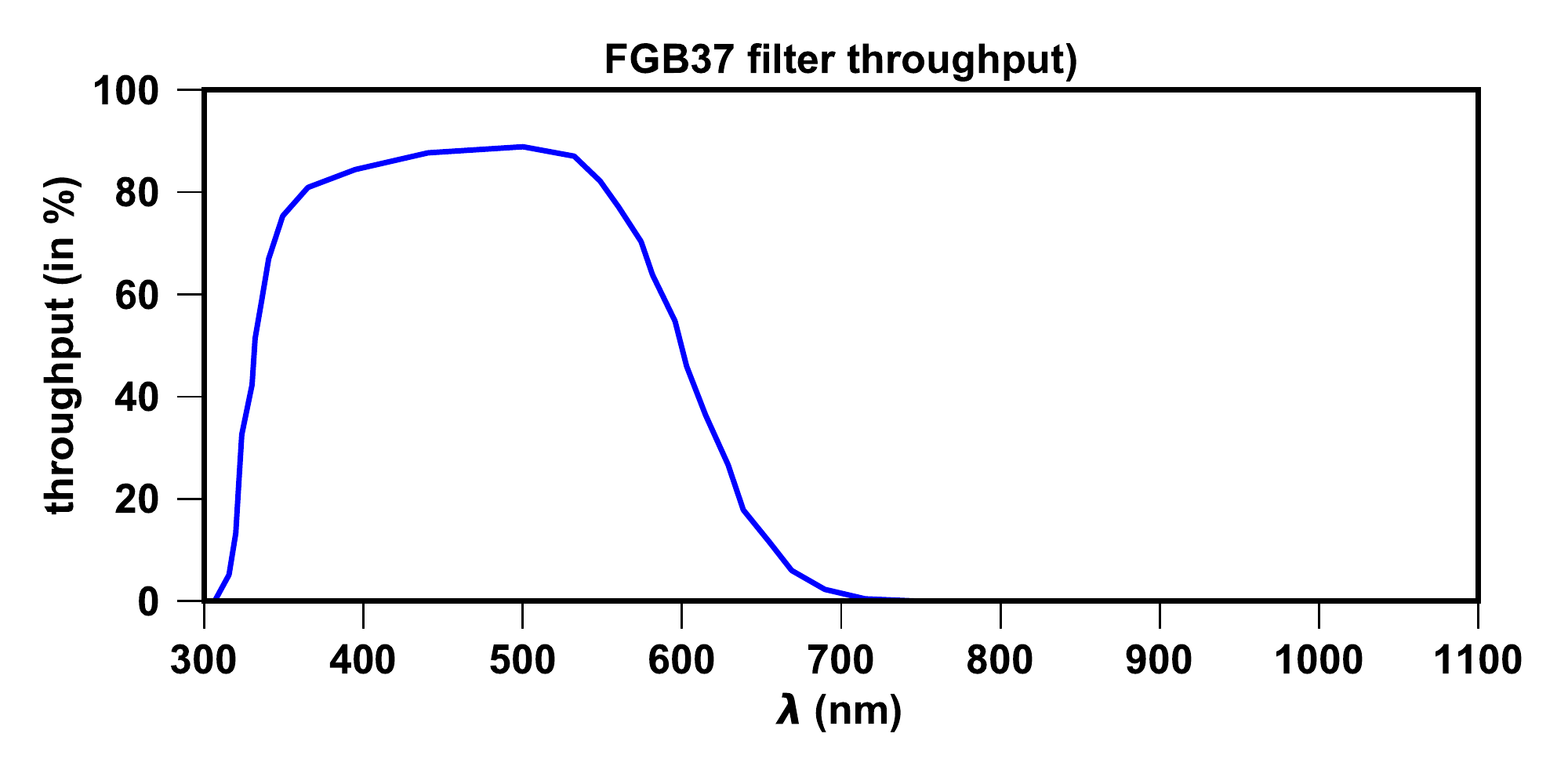}
\includegraphics[width=\columnwidth]{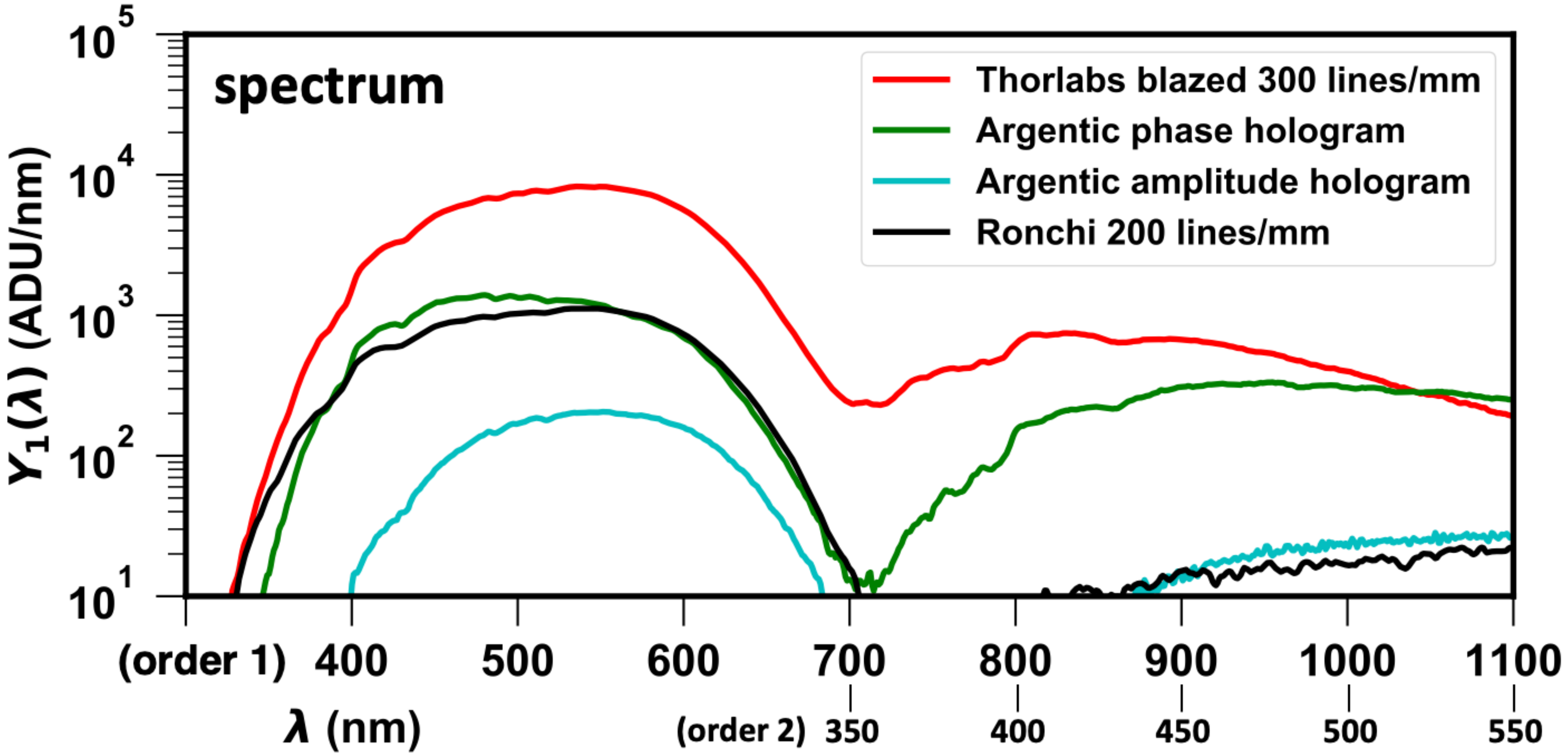}
\includegraphics[width=\columnwidth]{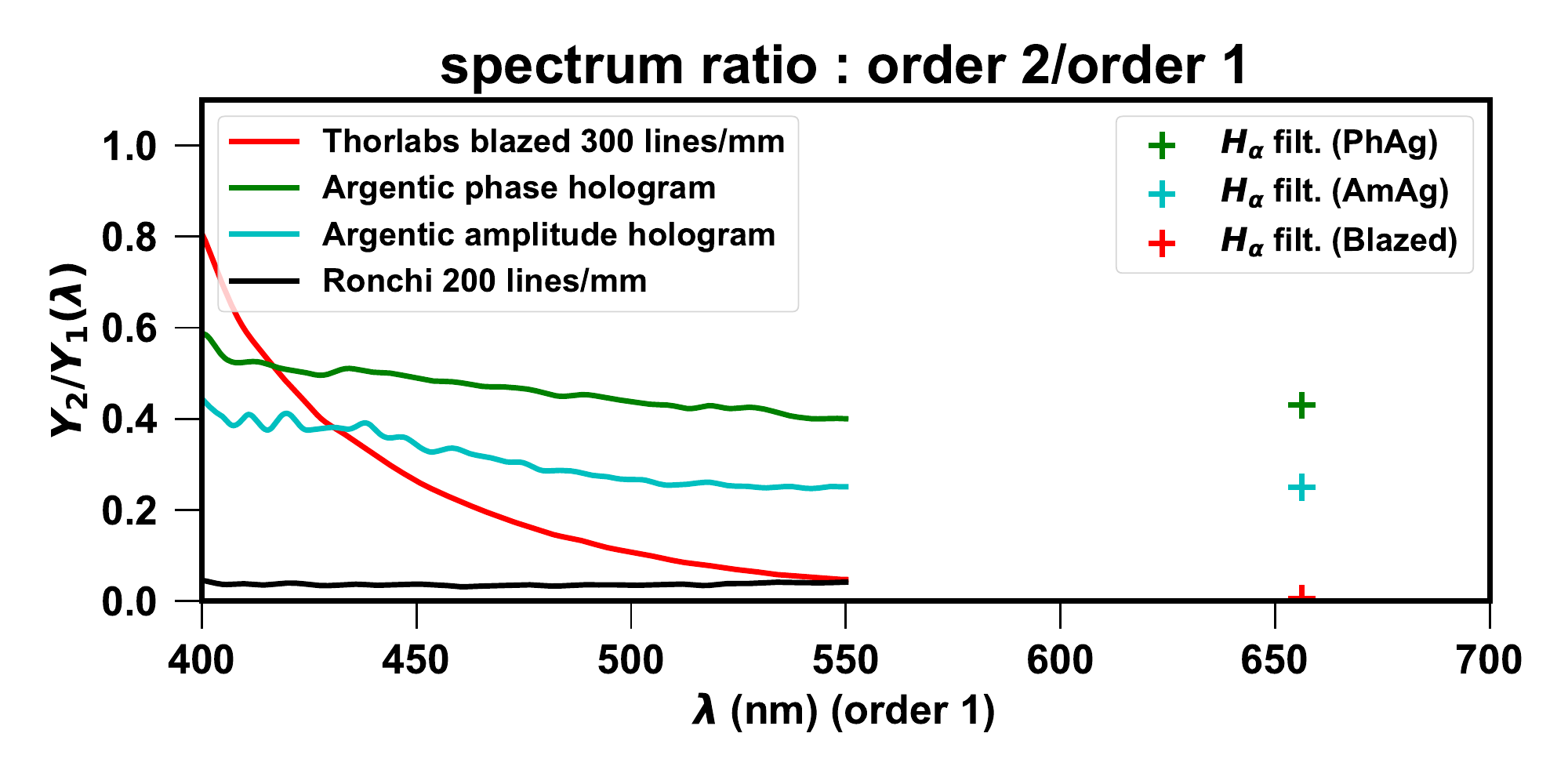}
\end{center}
\caption[] 
{Top: The FGB37 band-pass filter throughput.

Middle: the instrumental CALSPEC~HD205905 spectral density through the FGB37 blue band-pass filter, obtained with the Blazed300, a Ronchi grating with $200\,$lines/mm, and the holograms.
Here the fluxes from two orders of diffraction are superimposed, but thanks to the filter, the
first order (from 350 to 700\,nm) and second order (beyond 700\,nm, corresponding to $\lambda_{2nd\, order} > 350\,$nm), can clearly be distinguished.
The plotted values are flux densities per wavelength unit of the first diffraction order.
The second scale (order 2) shows the wavelength from the second order dispersion law.

Bottom: the measured second to first order transmission ratios.
}
\label{rapports}
\end{figure}

Since the spectral domain we want to study exceeds one octave, we expect superimposition of the blue part of the second diffraction order spectrum onto the red part of the first diffraction order.
Figure \ref{rapports} is produced from spectra obtained through the blue band-pass filter FGB37.
This filter absorbs all the light redder than $700\,$nm.
As a consequence, the blue light ($350<\lambda<550\,$nm) coming from the second order and reaching abscissa $700-1100$\,nm on the figure is the only contribution within this part of the experimental spectrum.
The lower panel of Fig. \ref{rapports} shows the deduced transmission ratio of the second to the first orders of diffraction, defined by the ratio of the light fluxes within $d\lambda$ that are diffracted in the second and first orders, {\it i.e.}:
\begin{equation}
Y_2/Y_1(\lambda)=\frac{dF_2(\lambda)}{d\lambda}\big/\frac{dF_1(\lambda)}{d\lambda}.
\end{equation}
We also estimated this ratio at $H_{\alpha}$ wavelength, by using the narrow band-pass $H_{\alpha}$ filter.
Thanks to the suppression of even orders in the Fourier transform of the square wave function characterizing the Ronchi grating (here we did the measurement with a $200\,$ lines/mm Ronchi grating instead of 400),
this type of disperser has the smallest second order relative contribution;
in the bluest part, the strongest relative second order is observed with the Blazed300. Amongst the holograms, the phase hologram shows the largest relative second order contribution, followed by the amplitude hologram.

\section{Discussion: toward a finalized hologram for AuxTel}
\label{Sect:discussion}

The prototypes studied here were our first generation of holograms, initially produced as a proof of concept to test their focusing properties. But we learned much more lessons from our tests and
we are now able to define more precisely the requirements for the final hologram to be installed on the AuxTel:
\begin{itemize}
    \item choose modulated phase hologram that has a significantly better transmission than the modulated amplitude hologram;
    \item maximise the light transmitted in the first (+1) diffraction order with the minimal objective to exceed the constant $10\%$ transmission of a Ronchi grating;
    \item minimise the light transmitted in the second order;
    \item minimise the transmission variation with the incoming beam position, by filtering the interfering laser beams on the holographic optical bench, and avoiding diffuse light during the hologram recording.
\end{itemize}
The transmission of the phase holograms can be improved by tuning different emulsion parameters (for instance the thickness of the emulsion or the size of the silver complex grains). Theoretically a maximum of $32\%$ of the light can be diffracted in the first order with a thin hologram \citep{Kogelnik_1969}, with reduced light in the second order.

An ideal, theoretical hologram should only produce +1 and -1 orders (reconstructed and conjugate object beams), in addition to the zero-th order.
However, from a theoretical hologram to a real hologram,  a non-linear intervening process (such as photographic recording) takes place. That introduces distortions in the record of the interference pattern, producing higher orders in the image restitution.
For the final hologram production, we explore a way to reduce the second order contribution by recording the interference pattern through very contrasted photographic emulsion, to mimic crenel phase variations.

Series of prototypes adapted to the AuxTel geometrical configuration have been produced to approach the theoretical best performances.
Their evaluation benefited from extensive optical test-bench measurements and will be described in a forthcoming paper \citep{Holospec2_2021}.

Since we cannot entirely suppress the second order, a complete analysis of the stellar spectrum will need to perform a combined fit of the first + second diffraction orders, assuming knowledge of the system throughput as a function of the wavelength.
An alternative strategy to cancel the second order (significant for $\lambda>700\,$nm) could be to use the hologram together with a low passband filter when only the red part of the spectrum has to be measured (for water vapor absorption estimates).
The choice of using or not such a filter for AuxTel observations could be synchronized with the choice of the filter used by the SST.

\section{Conclusion and perspectives}
\label{Sect:conclusion}
In this paper, we have tested the advantages of using a plane holographic grating as a disperser inserted on the path of a convergent telescope beam, to convert an imager telescope into an on-axis slitless spectrophotometer.
The spectrophotometer thus exclusively consists of the dispersive and focusing hologram and the sensor.

Systematic tests have been performed using the CTIO $0.9\,$m telescope equipped with an on-axis CCD camera.
First, we checked and quantified the benefits of the optical function of the hologram that ensure correct focusing for the complete spectrum from $370\,$nm to $1050\,$nm. Our tests have shown that the PSF of the first diffraction order does not significantly downgrades with respect to the PSF of the zero-th order. A significant improvement is specifically observed with respect to the periodic gratings in the red section of the spectra (for equivalent dispersions).
We determined the allowed excursion of the position of the source beam-axis around the optical center of the holograms in order to maintain optimal focusing performances. We found a "sweet pot" of at least $1\,$cm$^2$ area on the CCD plane, corresponding to an angular diameter of $1.5\,$arcmin on the sky.

As a conclusion, this kind of holographic grating can easily convert a telescope imaging camera into a slitless spectrograph with a rather good resolution at a moderate price. The main challenge to get accurately calibrated spectrum lies in the proper extraction of the spectrum from the image which will be detailed in \cite{Spectractor}.  
The only instrumental need is the availability of a free slot in a filter wheel.
A customized hologram has to be produced with a specific optical bench, only depending on the distance from the filter wheel to the camera CCD, and on the desired dispersive power.

\section*{acknowledgements}
We are grateful to the CTIO technical staff members
Hernan Tirado and Manuel Hernandez for their help during our tests with the CTIO $0.9\,$m telescope. We also thank M\'elanie Chevance for her participation to the observations and Augustin Guyonnet for fruitful advice for the CTIO image reduction. The cost of the observations have been shared by the IJCLab (IN2P3-CNRS) and the Department of Physics and Harvard-Smithsonian Center for Astrophysics, Harvard University.

This paper has undergone internal review in the LSST Dark Energy Science Collaboration. The internal reviewers were S. Bongard, Y. Copin and M. Coughlin.

Author contribution statement:
MM leads the hologram project from the idea to the paper writing. JN observed at CTIO, analyzed the data, characterized the holograms and produced their theoretical description. SDC analyzed the data, provided atmospheric expertise and performed optical simulations and their analysis. YG made the holographic optical elements. LLG supported the project. 

DESC acknowledges ongoing support from the IN2P3 (France), the STFC 
(United Kingdom), and the DOE, NSF, and LSST Corporation (United States).  
DESC uses resources of the IN2P3 Computing Center 
(CC-IN2P3--Lyon/Villeurbanne - France) funded by the Centre National de la
Recherche Scientifique; the National Energy Research Scientific Computing
Center, a DOE Office of Science User Facility supported under Contract 
No.\ DE-AC02-05CH11231; STFC DiRAC HPC Facilities, funded by UK BIS National 
E-infrastructure capital grants; and the UK particle physics grid, supported
by the GridPP Collaboration.  This work was performed in part under DOE 
Contract DE-AC02-76SF00515.

\section*{Data Availability}
The data underlying this article will be shared on reasonable request to the corresponding author.

\bibliographystyle{mnras}
\bibliography{citations_holospec} 

\appendix

\section{Simulation of the optical PSF and aberrations}
\label{sec:psfsim}
The structure of the beam spot at the focus of the CTIO $0.9\,$m telescope  ($f/13.7$, scale at focal plane $60\,\mu$m/arcsec)
can be assessed by optical simulation using Beamfour \citep{Beamfour_2016}.
Beamfour implements a standard ray-tracing procedure defined in ~\cite{Spencer:62}.
The trajectory of each optical ray is followed through the successive optical surfaces arranged along the optical axis of the experimental setup.
At each surface where the ray interacts, the new direction $\vec{S}^\prime$ is calculated from the previous direction $\vec{S}$ according the following equation:
\begin{equation}
n^\prime\vec{S}^\prime \wedge \vec{\zeta}= n \vec{S}  \wedge \vec{\zeta} + p \lambda N_{\rm eff} \vec{v},   
\label{eq:raytracingequation}
\end{equation}
where $\vec{S}$ and $\vec{S}^\prime$ are the unit vectors of the incident and outgoing ray, $n$ and $n^\prime$ are the refractive indexes of the medium upstream and downstream the optical surface, $p$ is the diffraction order and $N_{\rm eff}$ is the spatial density of the grating lines. 
The unit vectors direct base $(\vec{u},\vec{v},\vec{\zeta})$ is defined at the interaction point on the surface, where $\vec{\zeta}$ is the normal to the surface, $\vec{v}$ is parallel to the grating lines and $\vec{u}$ is perpendicular to the lines (similar to Fig. \ref{fig:frame} but on the hologram surface).
The second term to the right of equation~\ref{eq:raytracingequation} is relevant only for interfaces that have a grating with a defined orientation and local $N_{\rm eff}$ density of lines.
For our simulation we set $n=n^\prime = 1$, thus isolating the optical function of the disperser, ignoring its glass support. For the simulation of the hologram, the orientations and densities $N_{\rm eff}$ of the local grating are computed from the interference pattern of the two sources recorded on the hologram. For the simulation of the Ronchi disperser, the orientation and the value $N_{\rm eff}$ are fixed. 
Under these assumptions, and using $\vec{v}= - \vec{u}\wedge \vec{\zeta}$,
equation~\ref{eq:raytracingequation} becomes~:
\begin{equation}
  (\vec{S}^\prime - \vec{S} + \Lambda \vec{u}) \wedge \vec{\zeta}=\vec{0},
\end{equation}
where $\Lambda  =  p \lambda N_{\rm eff}$.
This means that vector $(\vec{S}^\prime - \vec{S} + \Lambda \vec{u})$ is collinear to $\vec{\zeta}$:
\begin{equation}
  \vec{S}^\prime - \vec{S} + \Lambda \vec{u} = \Gamma \vec{\zeta}
  \label{eq:raytracingdirection}
\end{equation}
where $\Gamma$, such that $\vec{S}^\prime$ norm is 1, must satisfy the following equation~:
\begin{equation}
\Gamma^2 + 2 (\vec{S}\cdot \vec{\zeta}) \Gamma + \Lambda^2 -2\Lambda (\vec{S}\cdot \vec{u})  =  0.
\label{eq:raytracinggamma}
\end{equation}
In our case of a transmission grating, the root with smallest module corresponds to the solution associated with the $p-$th order diffracted ray.
If there is no real root, then there is no diffracted ray at order $p$.

\begin{figure}
\begin{center}
\includegraphics[width=\columnwidth]{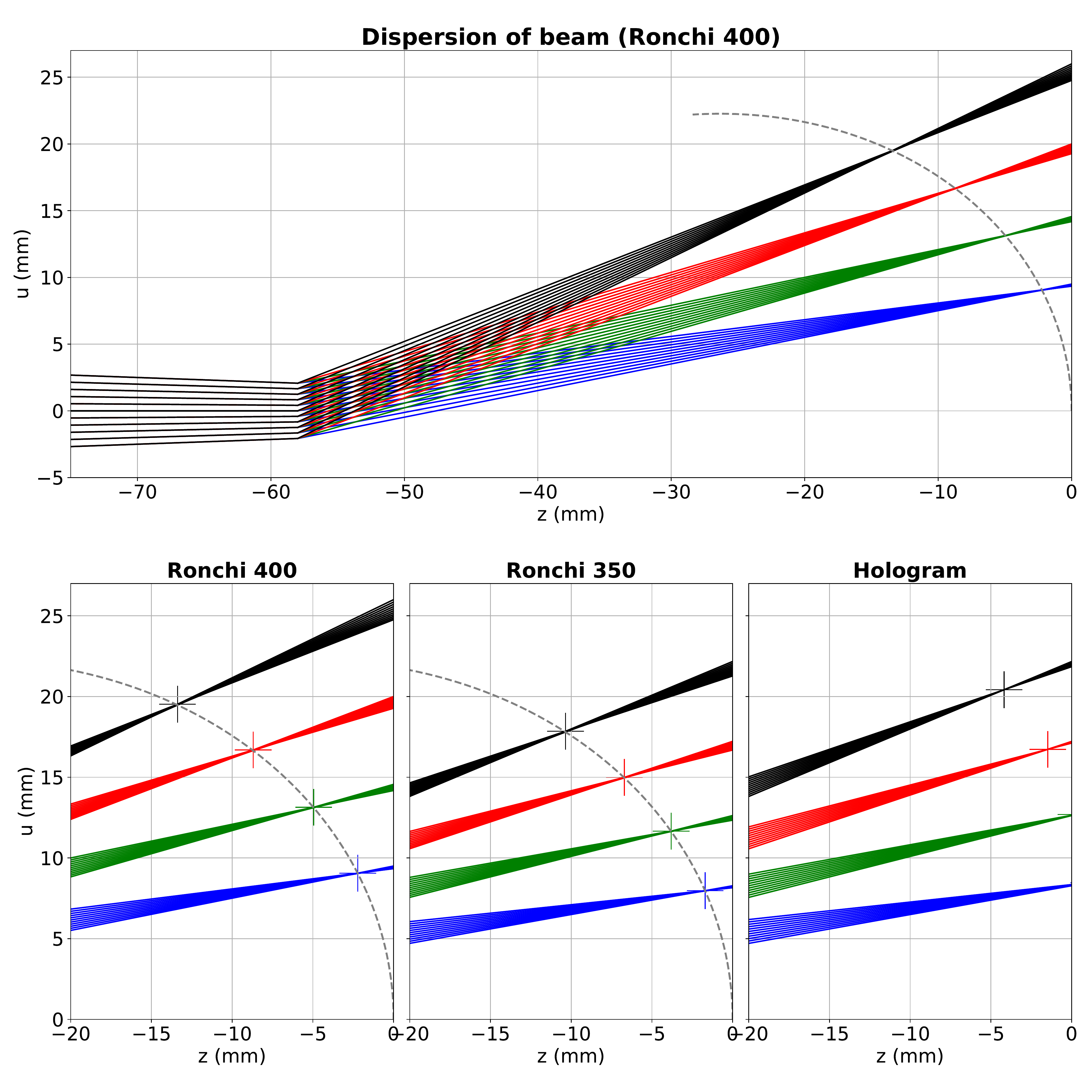}
\end{center}
\caption[] {
First diffraction order beam focus at 400, 600, 800, 1000~nm (blue, green, red, black), from a BEAMFOUR ray-tracing simulation in tangential plane. Top panel: beam dispersion for a Ronchi grating with 400 lines/mm. Bottom panels: zoom on defocusing near the sensor for Ronchi 400 lines/mm, Ronchi 350 lines/mm (for comparison with the hologram) and the hologram (with equivalent spatial frequency $\sim 350\,$lines/mm). Curves and crosses show the focus in the deflection plane, calculated from the grating equation.
}
\label{defocus}
\end{figure}

\begin{figure}
\begin{center}
\includegraphics[width=\columnwidth]{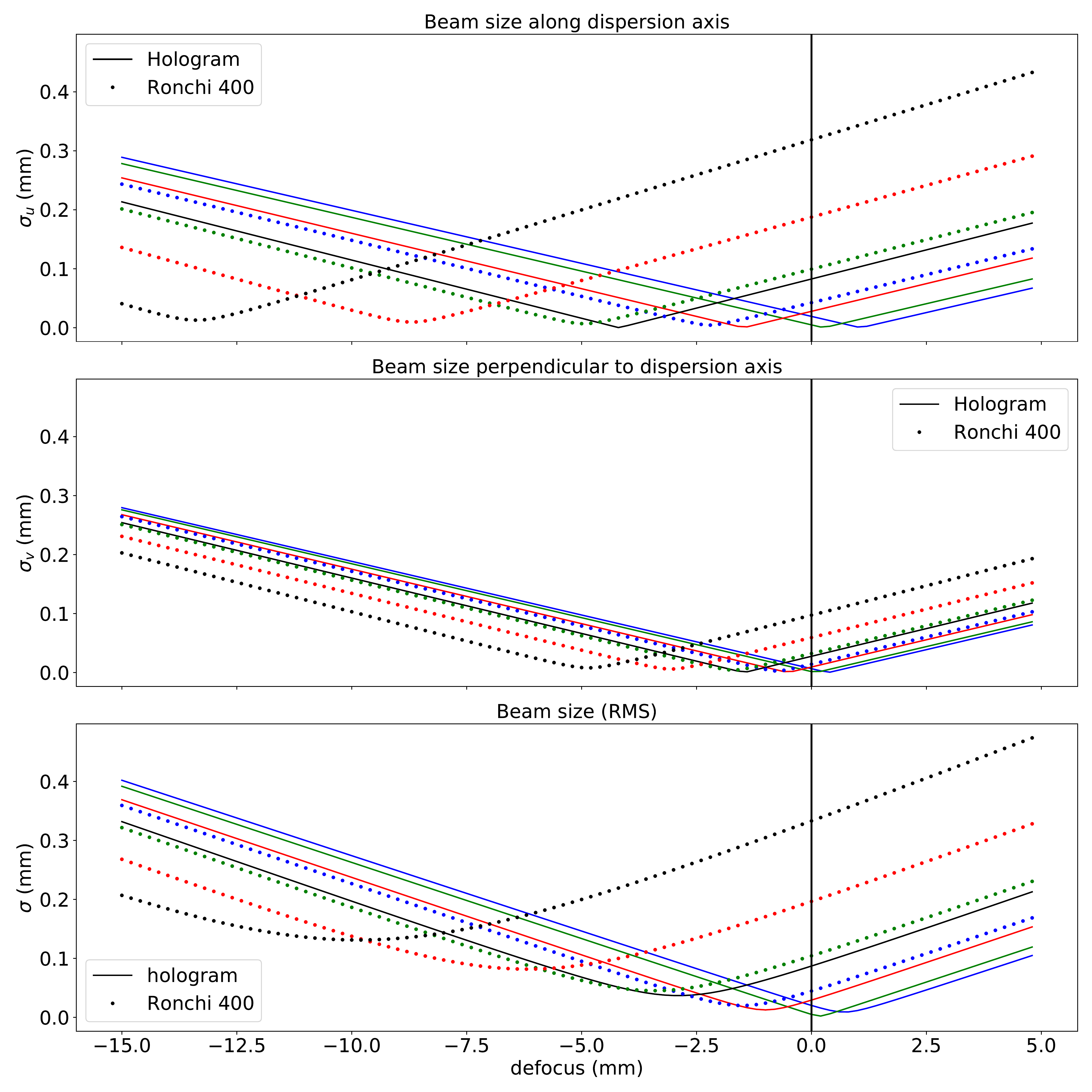}
\end{center}
\caption[] 
{\it
The beam spot size versus the shift from the focal plane of the zero order image for hologram and Ronchi400.
The vertical line shows the intersection of the focal plane.
Top panel : $\sigma_u$, along the dispersion axis,
middle panel : $\sigma_v$, perpendicular to the dispersion axis,
bottom panel : $\sigma=\sqrt{\sigma_u^2+\sigma_v^2}$.
Solid and dotted lines are used for hologram and Ronchi400.
Blue, green, red, black curves correspond to wavelengths 400,600, 800 and 1000$\,$nm.
}
\label{fig:defocus}
\end{figure}

We materialize the incoming beam cross-section with a disk shaped grid with light-rays uniformly distributed within the telescope aperture,
all converging at the focal point.
The disperser is inserted at distance $D_{\mathrm{CCD}}=58\,$mm upstream the focal point.
The beam profile is shown on Fig.~\ref{defocus} for 4 wavelengths, and its defocusing with respect to the zero order image is estimated from the beam spot size reported in Fig.~\ref{fig:defocus}.

\begin{figure}
\begin{center}
\begin{tabular}{c}
{\bf Hologram} \\
\includegraphics[width=\columnwidth]{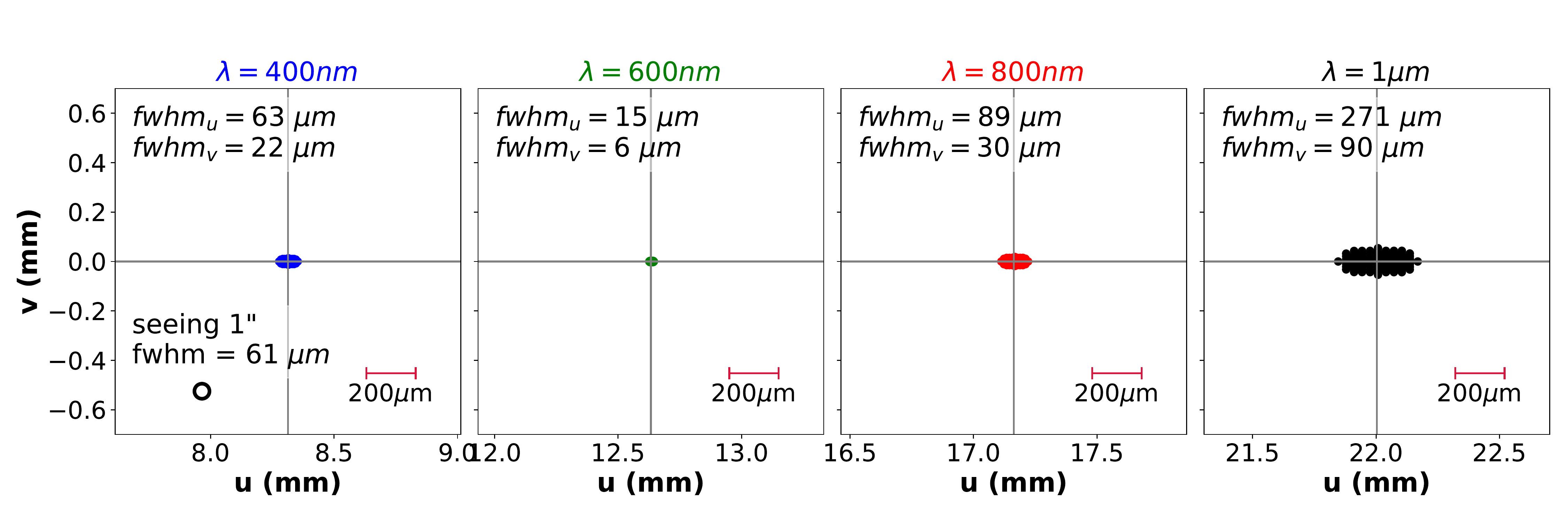} \\
{\bf Ronchi400} \\
\includegraphics[width=\columnwidth]{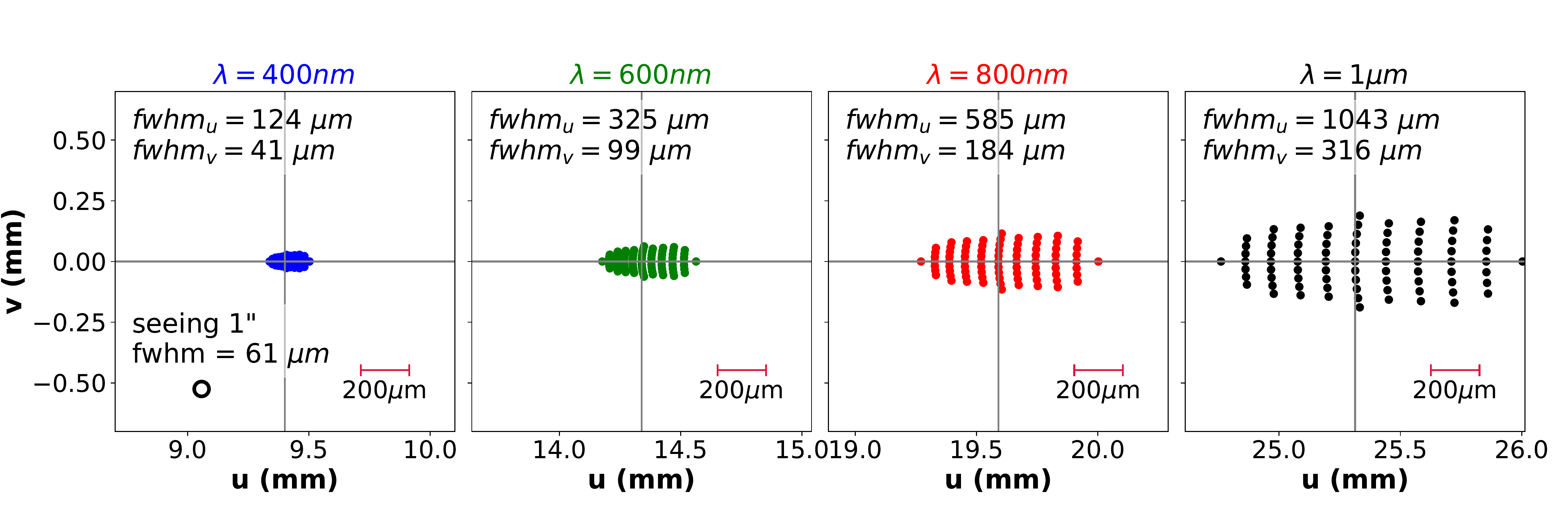}
\end{tabular}
\end{center}
\caption[] {Beam spot PSF in the focal plane for Hologram
(with $N_{\mathrm{eff}}$ corresponding to the position of the chief ray)
and Ronchi400 calculated with BEAMFOUR. The center of each panel is at the position predicted by the dispersion relation. The shift between the beam spot barycenter and the predicted position ranges between 2-5~$\mu$m for Hologram and 7-23~$\mu$m for Ronchi400, increasing slightly with wavelength; it remains always much smaller than the beam extension ${\rm fwhm}_u$. Panels for four wavelengths are shown for beam incident angle of $0\degree$.  The small circles on the lower-left corner show the extension of a 1 arcsec seeing spot.
}
\label{fig:psfsim}
\end{figure}

In agreement with simple geometric considerations from periodic gratings
(such as the Ronchi), it was expected that the larger the wavelength, the stronger the defocusing and aberration. This is clearly demonstrated in Fig. \ref{fig:defocus}.
Moreover, in the zero order focal plane, the beam spot size along the dispersion axis $\sigma_u$ is significantly larger than the size $\sigma_v$ perpendicular to the dispersion axis for the Ronchi400 case (Fig.~\ref{fig:psfsim}).
Unfortunately, $\sigma_u$ is the most critical parameter as it limits the spectroscopic resolution, since this aberration leads to wavelength mixing if the spectrum is naively evaluated from the perpendicular projection on the dispersion axis, without deconvolution.
Alternatively, by design, we expect no defocusing nor aberration for the hologram at its recording wavelength $\lambda_R=639\,$nm, which is indeed what Figs. \ref{fig:defocus} and \ref{fig:psfsim} show (green solid line curves and spot). Bluer wavelengths focus slightly downstream (blue solid line curve) whereas redder wavelengths focus slightly upstream,
due to next-order effects.
The important fact is that the hologram best focus ($\sigma_u$ waist) remains always within $4\,$mm of the focal plane whereas the Ronchi400 best focus is systematically upstream the focal plane (up to $13\,$mm).
For the hologram, this results in a beam spot size at the focal plane which is always comparable with the typical seeing spot expected at the telescope site (1 arcsec), except for $\lambda = 1000$~nm, for which $\sigma_u$ is slightly worse (see Fig.~\ref{fig:psfsim}).

\section{Consequences of the stationarity of $n(S'_0,P)$}\label{app:fermat}

As in~\cite{Noda:74}, a power-series expansion of $F(M)$ (expression (\ref{light-path})) in $u$, $v$ and $v_0$ gives :
\begin{align}
F(u,v) =& (v-v_0) \left[p\lambda \left.\frac{\partial n}{\partial v}\right\vert_{S'_0} \right] +  u \left[-\sin\theta_p+\sin\theta_0 + p\lambda \left.\frac{\partial n}{\partial u}\right\vert_{S'_0} \right]    \notag \\
& + \frac{u^2}{2} \left[\dfrac{\cos^2\theta_p}{r_p}-\dfrac{\cos^2\theta_0}{r_0}+ p\lambda \left.\frac{\partial^2 n}{\partial u^2}\right\vert_{S'_0} \right]  \notag \\
& + \frac{v^2-v_0^2}{2} \left[\dfrac{1}{r_p}-\dfrac{1}{r_0} + p\lambda \left.\frac{\partial^2 n}{\partial v^2}\right\vert_{S'_0} \right]  \notag  + {u v}\left[ p\lambda \left.\frac{\partial^2 n}{\partial u\partial v}\right\vert_{S'_0}  \right] \notag \\
& + \text{higher order terms...}
\end{align}
where $r_p\sin\theta_p=u_p(\lambda)$ and $r_0\sin\theta_0=u_0$. To compute the position of the image $S_p(\lambda)$, the stationarity condition of the Fermat's principle states that each of these terms must cancel.

Therefore, the cancellation of the first term of the $F(M)$ expansion implies that the dispersion axis is locally orthogonal to the mean orientation of the grooves at $S'_0$ and follows the gradient of the interference pattern recorded on the hologram. In the $(w,l,\zeta)$ frame, the orientation of the dispersion axis is the angle $\alpha(w'_0,l'_0)$ of the dispersion axis:
\begin{equation}\label{eq:alpha}
\tan\left(\alpha(w'_0,l'_0) - \alpha_0\right) = \dfrac{(\left.\partial n / \partial l) \right\vert_{S'_0}}{\left.(\partial n / \partial w) \right\vert_{S'_0}}
\end{equation}
with $\alpha_0$ the mean rotation of the grating expressed in the sensor frame. It means also that if $\left.(\partial n / \partial l) \right\vert_{S'_0} = 0$ then the $u$ axis is aligned with the $w$ axis. This is the case only if $S'_0$ is on the $w$ axis.

The second term gives the grating formula for transmission holograms along the $u$ axis with $S'_0$ being the incident point of the chief ray :
\begin{equation}\label{eq:dispersion_fermat}
\sin\theta_p(\lambda)-\sin\theta_0 = p\lambda \left.\frac{\partial n}{\partial u}\right\vert_{S'_0}
\end{equation}
and the map of the effective groove density of the hologram is:
\begin{equation}\label{eq:Neff}
N_{\mathrm{eff}}(w'_0,l'_0) = \left.\frac{\partial n}{\partial u}\right\vert_{S'_0} = \sqrt{1+  \left(\dfrac{(\left.\partial n / \partial l) \right\vert_{S'_0}}{\left.(\partial n / \partial w) \right\vert_{S'_0}}\right)^2} \left.\frac{\partial n}{\partial w}\right\vert_{S'_0}
\end{equation}
given that $\left.(\partial n / \partial v) \right\vert_{S'_0} = 0$.
The grating formula for holographic gratings used in convergent beam is thus the same as the usual grating formula for regular disperser with a local $N_{\mathrm{eff}}(w'_0,l'_0)$ line density.

The cancellation of the third term gives the tangential focus $f'_t(S'_0)$:
\begin{equation}
\dfrac{\cos^2\theta_p}{f'_t}-\dfrac{\cos^2\theta_0}{r_0}+ p\lambda \left.\frac{\partial^2 n}{\partial u^2}\right\vert_{S'_0}=0
\end{equation}
\begin{equation}
\Rightarrow f'_t(w'_0,l'_0,\lambda) = \dfrac{D_{\mathrm{CCD}} \cos^2\theta_p(\lambda)}{\cos^3\theta_0 -  p\lambda  D_{\mathrm{CCD}}\left.\dfrac{\partial^2 n}{\partial u^2}\right\vert_{S'_0}}
\end{equation}
with $r_0 = D_\mathrm{CCD} / \cos\theta_0$ if the zero-th order is correctly focused on the CCD. At last, the fourth term leads to the sagittal focus  $f'_s(S'_0)$:
\begin{equation}
\dfrac{1}{f'_s}-\dfrac{1}{r_0}+ p\lambda  \left.\frac{\partial^2 n}{\partial v^2}\right\vert_{S'_0}=0
\end{equation}
\begin{equation}
\Rightarrow f'_s(w'_0,l'_0,\lambda) = \dfrac{D_{\mathrm{CCD}}}{\cos\theta_0 -  p\lambda  D_{\mathrm{CCD}} \left.\dfrac{\partial^2 n}{\partial v^2}\right\vert_{S'_0}}
\end{equation}
This term models the bending of the spectrum around the dispersion axis defined by $\left.(\partial n / \partial l) \right\vert_{S'_0} = 0$.

The higher order terms in the $F(M)$ expansion are related to the coma aberrations; they are cancelled or greatly reduced by the hologram conception \citep{Murty:62}. Note again that all these results are included in \cite{Noda:74} and \cite{Palmer:89}, but we reformulate them here for the particular case of a planar holographic grating to be used directly in a slitless spectrograph.

For completeness, we write here below the first derivatives of the $n(S'_0,P)$ function, that are needed to compute the groove density $N_\mathrm{eff}(w'_0,l'_0)$ of holograms, as well as the dispersion axis angle and the focal curves.

Given that, in the $(w,l,\zeta)$ frame, the source positions are $A(-d/2,0,D_R^{(A)})$ and $B(d/2,0,D_R^{(B)})$, we get:
\begin{align}
 \left.\frac{\partial n}{\partial w}\right\vert_{S'_0} = \frac{1}{\lambda_R} &\left[  \frac{d/2-w'_0}{\sqrt{(d/2-w'_0)^2+(l'_0)^2+(D_R^{(B)})^2}} \right. \notag \\
 & \left.  - \frac{-d/2-w'_0}{\sqrt{(-d/2-w'_0)^2+(l'_0)^2+(D_R^{(A)})^2}} \right]
\end{align}

\begin{align}
 \left.\frac{\partial n}{\partial l}\right\vert_{S'_0} = \frac{1}{\lambda_R} &\left[  -\frac{l'_0}{\sqrt{(d/2-w'_0)^2+(l'_0)^2+(D_R^{(B)})^2}}\right. \notag \\
 & \left.  +\frac{l'_0}{\sqrt{(-d/2-w'_0)^2+(l'_0)^2+(D_R^{(A)})^2}} \right].
\end{align}
In the $(u,v,z)$ frame the source positions $A(u_A,v_A,z_A)$ and $B(u_B,v_B,z_B)$ are given by:
\begin{equation}
    \left\lbrace \begin{array}{ll}
    u_i = &\ \left(w_i-w'_0\right)\cos \beta  - l'_0 \sin \beta \\
    v_i = & -\left(w_i-w'_0\right)\sin \beta  - l'_0\cos \beta  \\
    z_i = & D_\mathrm{CCD} - \zeta_i
    \end{array}\right.
\end{equation}
for $i=A,B$ and $\tan \beta = {(\left.\partial n / \partial l) \right\vert_{S'_0}}/{\left.(\partial n / \partial w) \right\vert_{S'_0}}$. The derivatives are then:

\begin{align}
 \left.\frac{\partial^2 n}{\partial u^2}\right\vert_{S'_0} = \frac{1}{\lambda_R} &\left[  -\frac{z_B^2+v_B^2}{\left(u_B^2+v_B^2+z_B^2\right)^{3/2}}  +\frac{z_A^2+v_A^2}{\left(u_A^2+v_A^2+z_A^2\right)^{3/2}} \right]
\end{align}

\begin{align}
 \left.\frac{\partial^2 n}{\partial v^2}\right\vert_{S'_0} = \frac{1}{\lambda_R} &\left[  -\frac{z_B^2+u_B^2}{\left(u_B^2+v_B^2+z_B^2\right)^{3/2}}  +\frac{z_A^2+u_A^2}{\left(u_A^2+v_A^2+z_A^2\right)^{3/2}} \right]
\end{align}

\section{Hessian analysis}\label{appendix-hessian}

The spectra have filament shapes that can be detected using an Hessian analysis inspired by the one developed in \citet{PlanckFilament}. The advantage of this technique is that it provides an analytical expression of the angle of the detected shape with respect to the horizontal or vertical axis of the CCD grid.

The Hessian matrix $H(x,y)$ is computed for each $x,y$ pixel of the image $I(x,y)$ as:
\begin{equation}
\displaystyle    {H(x,y) = \begin{pmatrix}
    H_{xx} & H_{xy} \\
    H_{xy} & H_{yy}
    \end{pmatrix} = \begin{pmatrix}
    \cfrac{\partial^2 I}{\partial x^2} & \cfrac{\partial^2 I}{\partial x \partial y} \\
    \cfrac{\partial^2 I}{\partial x \partial y} &  \cfrac{\partial^2 I}{\partial y^2} 
    \end{pmatrix} }
\end{equation}
The two eigenvalues of the Hessian matrix $H$ are calculated as
\begin{equation}
    \lambda_{\pm}(x,y) = \frac{1}{2} \left(H_{xx} + H_{yy} \pm h  \right)
\end{equation}
with $h = \sqrt{(H_{xx}-H_{yy})^2 + 4H_{xy}^2}$.
The value $\lambda_-$ corresponds to the eigenvector along the spectrum main axis while $\lambda_+$ corresponds to the eigenvector along the line of greater slope in intensity {\it i.e.} transverse to the dispersion axis. The orientation of these vectors can be analytically computed, for instance for $\lambda_-$ we find:
\begin{equation}
    \alpha_-(x,y) = \arctan \left(\frac{H_{yy} - H_{xx} - h}{2 H_{xy}}  \right) = \frac{1}{2} \arctan \left( \frac{2 H_{xy}}{H_{xx} - H_{yy}}  \right) 
\end{equation}
using the trigonometric formula $\tan 2\alpha = 2 \tan\alpha / (1- \tan^2 \alpha)$. After selecting all the pixels with $\lambda_-$ value above a reasonable threshold, the median $\alpha$ of the remaining $\alpha_-(x,y)$ values gives the mean orientation of the spectrum with respect to the $x$ axis. A linear fit can also be performed across the selected pixels and the slope gives an angle very similar to the one estimated with the median of the angle values.

\section{Phase hologram characterisation}\label{appendix_phase}

\begin{figure}
\begin{center}
\includegraphics[width=\columnwidth]{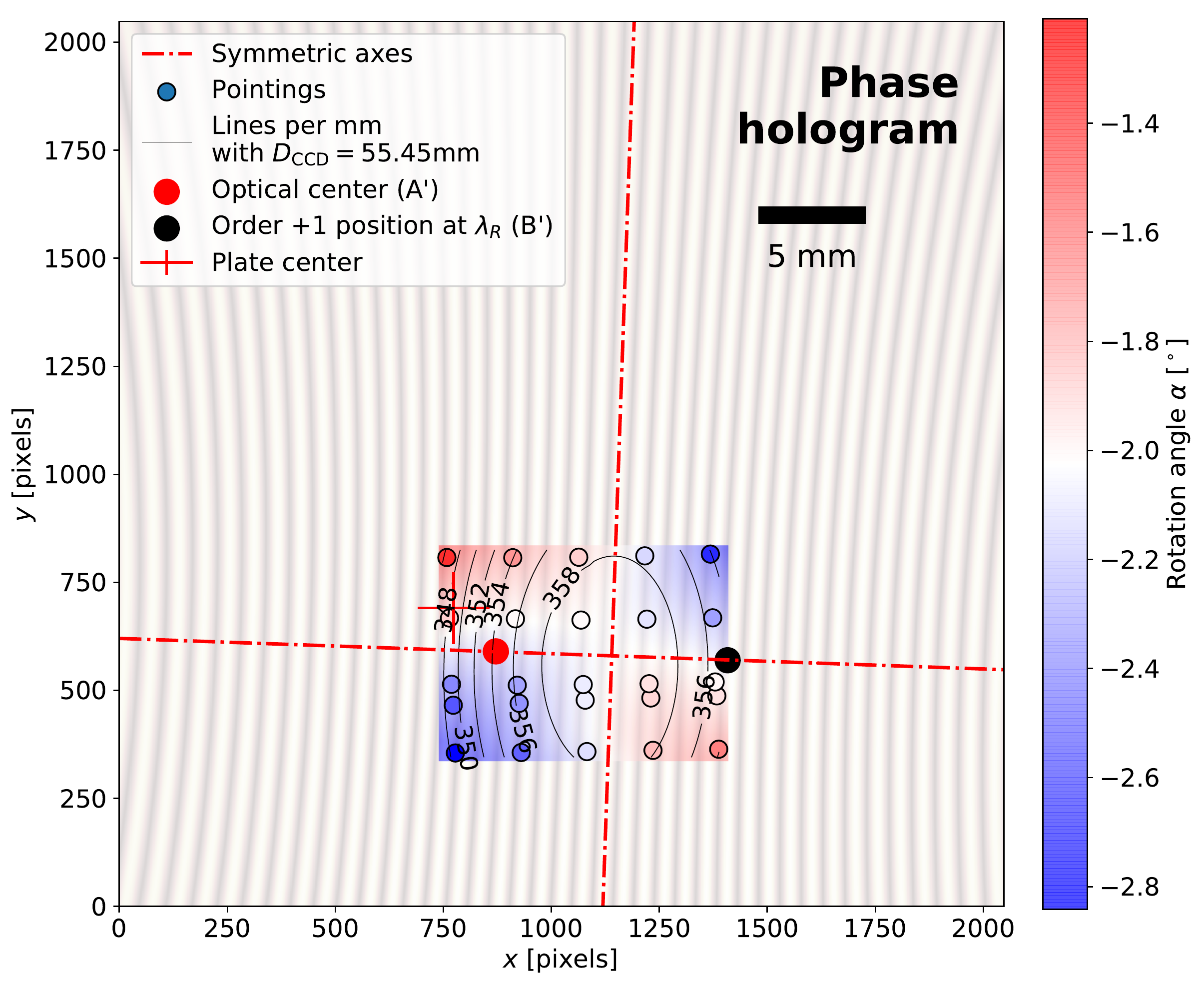}
\end{center}
\caption[] 
{Same as Figure~\ref{geometrie-holo} but for the phase hologram.}
\label{geometrie-holo-phag}
\end{figure}

\begin{figure}
\begin{center}
\includegraphics[width=0.8\columnwidth]{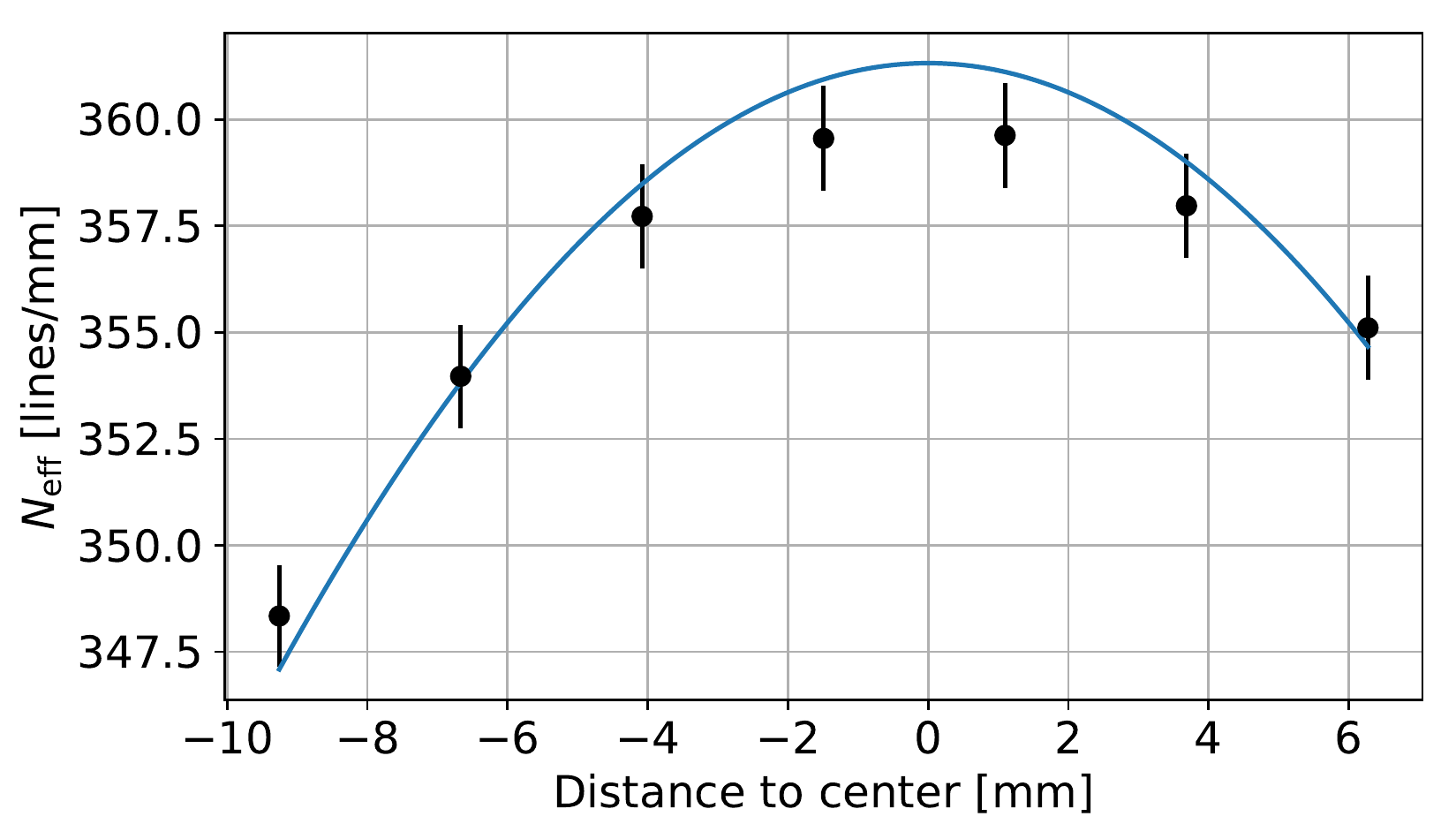}
\end{center}
\caption[] 
{Same as Figure~\ref{geometrie-holo-Neff} but for the phase hologram.}
\label{geometrie-holo-Neff-phag}
\end{figure}

\begin{figure}
\begin{center}
\includegraphics[width=\columnwidth]{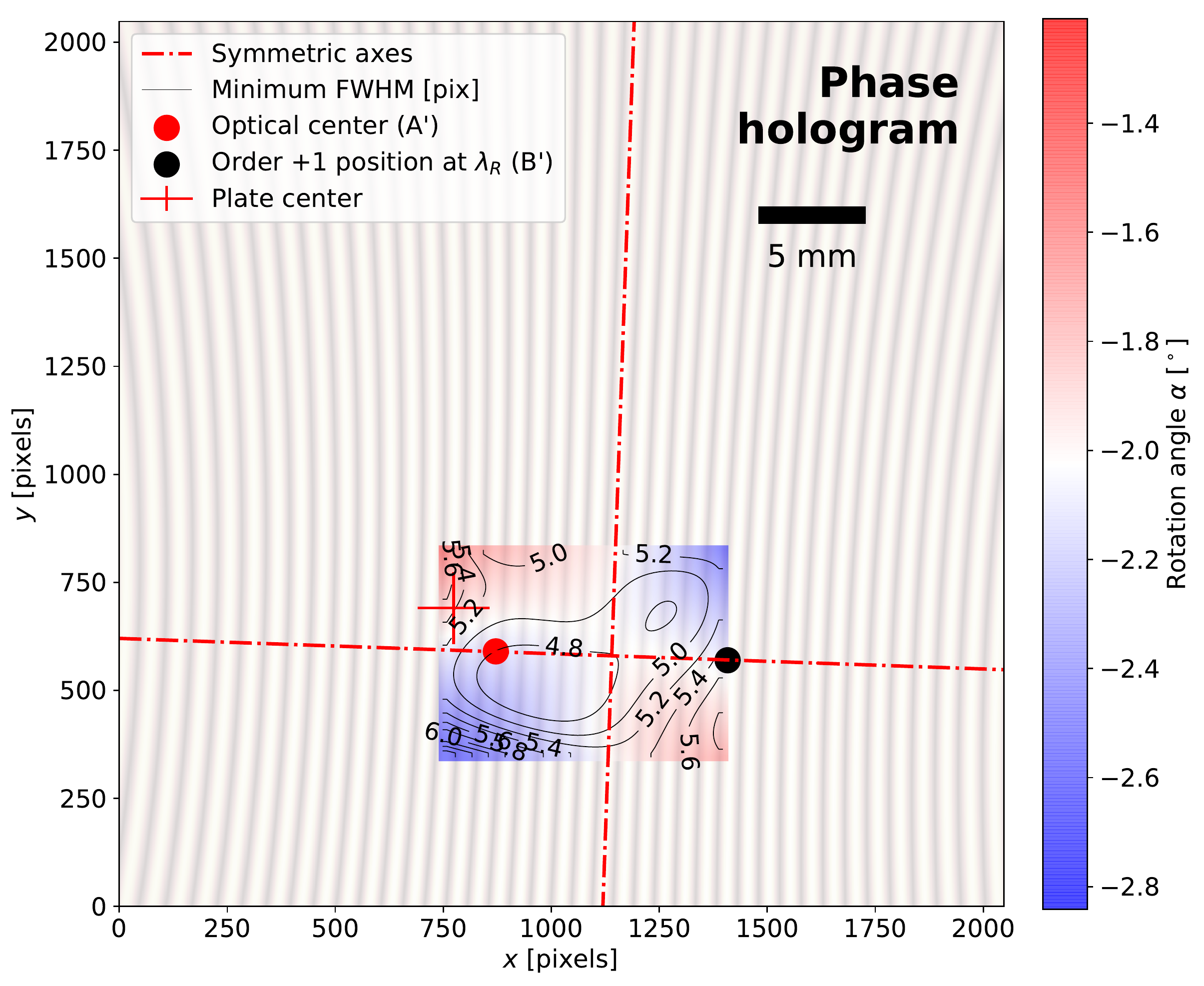}
\end{center}
\caption[] 
{Same as Figure~\ref{geometrie-holo-fwhm} but for the phase hologram.}
\label{geometrie-holo-fwhm-phag}
\end{figure}

The phase hologram has also been studied extensively and we present the results in Figures~\ref{geometrie-holo-phag},~\ref{geometrie-holo-Neff-phag} and \ref{geometrie-holo-fwhm-phag}.

\bsp
\label{lastpage}
\end{document}